%% file: main.tex
\documentclass[a4paper,11pt]{article}
\usepackage[mathlines]{lineno}
\usepackage{jcappub}

\usepackage[table,svgnames,dvipsnames]{xcolor}

\usepackage[normalem]{ulem}
\usepackage{aas_macros}
\usepackage{subfigure}
\usepackage{graphicx}% Include figure files
\usepackage{graphics}
\usepackage{dcolumn}% Align table columns on decimal point
\usepackage{bm}% bold math
\usepackage{orcidlink}
\usepackage{cases}% for dealing with mathematics,
\usepackage{booktabs}
\usepackage{comment}
\usepackage{multirow}
\usepackage{makecell}
\usepackage{siunitx}
\usepackage{tabularx}
\usepackage{xspace}
\usepackage{soul} % can comment out at the end
\usepackage{tablefootnote}
\graphicspath{{figs/}}

\usepackage{orcidlink} % for author list
\usepackage{hanging} % for affiliations
\usepackage{arydshln} % for horizontal dashed line, \hdashline
\usepackage{makecell} % for splitting cell in table into multiple rows
\newcommand{\cwr}[1]{{#1}}
\newcommand{\cwrminor}[1]{{#1}}
\newcommand{\refedit}[1]{\textcolor{black}{#1}}  %
\newcommand{\refeditt}[1]{\textcolor{black}{#1}}  % Updates after referee report
\newcommand{\figref}[1]{Fig.~\ref{fig:#1}}
\newcommand{\tabref}[1]{Table~\ref{table:#1}}
\newcommand{\secref}[1]{Section~\ref{sec:#1}}
\def\be{\begin{equation}}
\def\ee{\end{equation}}

\def\ba#1\ea{\begin{align*}#1\end{align*}}

\renewcommand{\emph}[1]{\textit{#1}}
\definecolor{RoyalBlue}{rgb}{0.25,.41,.88}
\definecolor{WildStrawberry}{HTML}{EE2967}
\definecolor{RedWine}{rgb}{0.743,0,0}
\definecolor{bittersweet}{rgb}{1.0, 0.44, 0.37}
\definecolor{burntorange}{rgb}{0.8, 0.33, 0.0}
\definecolor{midnightgreen}{rgb}{0.0, 0.29, 0.33}
\definecolor{otherblue}{rgb}{0.20, 0.73, 0.92}

\usepackage[nameinlink,noabbrev]{cleveref}
\crefname{equation}{Eq.}{Eqs.}
\crefname{section}{Section}{Sections}
\crefname{figure}{Figure}{Figures}
\crefname{table}{Table}{Tables}
\crefname{appendix}{Appendix}{Appendices}
\Crefname{figure}{Figure}{Figures}
\Crefname{equation}{Equation}{Equations}
\Crefname{section}{Section}{Sections}
\Crefname{table}{Table}{Tables}

\newcommand{\mksym}[1]{\ifmmode {\rm #1}\else #1\fi}

\renewcommand{\P}{\mathcal{P}}

\newcommand{\Om}{\Omega_\mathrm{m}}
\newcommand{\Ocdm}{\Omega_\mathrm{cdm}}
\newcommand{\Ob}{\Omega_\mathrm{b}}

\newcommand{\Or}{\Omega_\mathrm{R}}
\newcommand{\OL}{\Omega_\Lambda}
\newcommand{\ob}{\omega_\mathrm{b}}

\newcommand{\obc}{\omega_\mathrm{bc}}
\newcommand{\Obc}{\Omega_\mathrm{bc}}
\newcommand{\ocdm}{\omega_\mathrm{cdm}}

\newcommand{\Neff}{N_{\mathrm{eff}}}

\newcommand{\lcdm}{$\Lambda$CDM\xspace}

\newcommand{\lya}{Ly$\alpha$\xspace}

\newcommand{\DVrd}{D_\mathrm{V}/r_\mathrm{d}}
\newcommand{\DMrd}{D_\mathrm{M}/r_\mathrm{d}}
\newcommand{\DHrd}{D_\mathrm{H}/r_\mathrm{d}}
\newcommand{\DM}{D_\mathrm{M}}
\newcommand{\DV}{D_\mathrm{V}}

\newcommand{\rd}{r_\mathrm{d}}

\newcommand{\cs}{c_\mathrm{s}}

\newcommand{\sumnu}{\sum m_\nu}

\newcommand{\Planck}{\emph{Planck}\xspace}

\newcommand{\rs}{r_{\mathrm{s}}}
\newcommand{\keq}{k_{\mathrm{eq}}}
\newcommand{\logA}{\ln{\left(10^{10} 
A_\mathrm{s}\right)}}
\newcommand{\ns}{n_{\mathrm{s}}}
\newcommand{\qbao}{q_{\mathrm{BAO}}}
\newcommand{\tstar}{\theta_*}

\newcommand{\hinvmpc}{\,h^{-1}{\rm Mpc}}
\newcommand{\hmpcinv}{\,h\,{\rm Mpc^{-1}}}
\newcommand{\kmsMpc}{\,{\rm km\,s^{-1}\,Mpc^{-1}}}

\newcommand{\planckact}{\emph{Planck}/ACT\xspace}
\newcommand{\dessixcrosstwopt}{DES Y3 6$\times$2pt\xspace}
\newcommand{\actcrossunwise}{\emph{Planck}/ACT$\times$unWISE\xspace}

\title{$H_0$ Without the Sound Horizon (or Supernovae): A 2\% Measurement in DESI DR1}

\input{DESI-2025-0540_authors_2025.10.21}

\emailAdd{zaborowski.11@osu.edu}

\date{\today}

\abstract{
The sound horizon scale $\rs$ is a key source of information for \refedit{measurements of $H_0$ from early-time data}, and is therefore a common target of new physics proposed to solve the Hubble tension.
We present a sub-2\% measurement of the Hubble constant that is independent of this scale, using data from the first data release of the Dark Energy Spectroscopic Instrument (DESI DR1).
Building on previous work, we remove dependency on the sound horizon size using a heuristic rescaling procedure at the power spectrum level. 
A key innovation is the inclusion of \emph{uncalibrated} (agnostic to $\rs$) post-reconstruction BAO measurements from DESI DR1, as well as using the CMB acoustic scale $\tstar$ as a high-redshift anchor.
Uncalibrated type-Ia supernovae are often included as an independent source of $\Om$ information; here we demonstrate the robustness of our results by additionally considering two supernova-independent alternative datasets. 
We find somewhat higher values of $H_0$ relative to our previous work: $69.2^{+1.3}_{-1.4}$, $70.3^{+1.4}_{-1.2}$, and \cwr{$69.6^{+1.3}_{-1.8}\,\kmsMpc$} respectively when including measurements from i) \emph{Planck}/ACT CMB lensing $\times$ unWISE galaxies, ii) the DES Year 3 6$\times$2pt analysis, and iii) \emph{Planck}/ACT CMB lensing + the DES Year 5 supernova analysis.
These remarkably consistent constraints achieve better than 2\% precision; they are \cwrminor{among} the most stringent sound horizon-independent measurements from LSS to date, and provide a powerful avenue for probing the origin of the Hubble tension.
}

\begin{document}
\maketitle
\flushbottom

%%%%%%%%%%%%%%%%%%%%%%%%%%%%%%%%%%%%%%%%%%%%%%%%%%%%%%%%%%%%
\section{Introduction}
\label{sec:intro}
%%%%%%%%%%%%%%%%%%%%%%%%%%%%%%%%%%%%%%%%%%%%%%%%%%%%%%%%%%%%

The Hubble tension \cite{Verde:2019,DiValentino:2021a,Abdalla:2022,DiValentino:2021b}, recently reaching as high as the $6.4\sigma$ level \cite{Camphuis:2025}, continues to present one of the most significant challenges to the standard model of cosmology, \lcdm.
As additional data continues to reinforce this discrepancy in the local value of the expansion rate of the Universe, denoted $H_0$, between late-time probes (e.g. type-Ia supernovae, or SNe Ia) and early-time probes (e.g. the cosmic microwave background, or CMB), it is becoming increasingly likely that the resolution will lie either in undiscovered systematics or, enticingly, in modifications to our current understanding of physics \cite{DiValentino:2021a} (also see e.g. \cite{Wojtak:2022,Wojtak:2024,Christa:2024} for more recent investigations of systematics).
Many new theories of physics have therefore been proposed to ameliorate the Hubble tension \cite{Knox:2020}.
A common target of many such theories is to modify the predicted size of the sound horizon at recombination\footnote{We add the usual disclaimer that there is a subtle difference between the sound horizon scale at photon decoupling ($\rs$), which determines e.g. the observed power spectrum of the CMB, and the sound horizon scale at the so-called baryon drag epoch ($\rd$), which is relevant for BAO analyses and happens at a slightly later time because of the sub-dominance of baryons relative to photons. Where confusion is not expected to arise, we may refer to either scale as the sound horizon.}, because many of the most powerful early-time cosmological probes, such as the CMB and baryon acoustic oscillations (BAO), rely on this scale as their primary source of $H_0$ information.
Often called simply the ``sound horizon'', this scale (once calibrated) is a standard ruler that can be measured in these datasets and used as a probe of the Universe's expansion history (see \secref{theory}). 
Some of the proposed mechanisms to modify the sound horizon scale include, for example, early dark energy (EDE) \cite{Doran:2006,Bielefeld:2013,Karwal:2016}, extra relativistic degrees of freedom \cite{Eisenstein:2004,Hou:2013}, and exotic interactions in the neutrino sector \cite{Cyr-Racine:2014,Lancaster:2017,Kreisch:2020}.

Motivated by this fact, in \cite{Zaborowski:2025} we performed a sound horizon-free measurement of $H_0$ \refeditt{in the \lcdm model}, using data from the first data release of the Dark Energy Spectroscopic Instrument (DESI DR1; \cite{Snowmass2013.Levi, DESI2024.I.DR1}).
\refeditt{Conceptually, this (model-dependent) technique may be thought of both as an independent probe of the Hubble tension (i.e., ``is the sound horizon-independent $H_0$ constraint also in tension with local distance ladder determinations?''), as well as a cosmological null-test (``are the sound horizon-dependent and sound horizon-independent $H_0$ constraints consistent with each other?'').}
Combining DESI DR1 with measurements of CMB lensing from the ACT \cite{Das:2011,Sherwin:2017} and \Planck \cite{Planck:2014, Planck:2016lens,Planck:2020lens, Carron:2022} surveys, and separately including type-Ia supernova data from each of the 
Dark Energy Survey Year 5 (DES Y5 SN; \cite{DES:2024sne}), Pantheon+ \cite{Scolnic:2022}, and Union3 \cite{Rubin:2023} supernova analyses, resulted in sound horizon-free $H_0$ constraints of $66.7^{+1.7}_{-1.9}$, $67.9^{+1.9}_{-2.1}$, and $67.8^{+2.0}_{-2.2} \kmsMpc$, respectively. 
In that work, we used the heuristic sound horizon rescaling procedure originally developed in \cite{Farren:2022} and later deployed in \cite{Philcox:2022} (both using data from the BOSS survey \cite{Dawson:2012, Beutler:2016}) in order to marginalize over the sound horizon information contained in the galaxy power spectrum.
We review this procedure in \secref{theory} \refeditt{and the conclusions of our previous work in \secref{conclusions}}.

Our measurement in \cite{Zaborowski:2025} represented one of the tightest sound horizon-free $H_0$ constraints from large-scale structure data at the time\footnote{We note that in \cite{Brieden:2022} the authors achieved a similarly precise constraint using a template-based approach applied to data from the BOSS survey.}, amid a growing wave of interest in such techniques \cite{Baxter:2020,Philcox:2021,Brieden:2023,Smith:2023,Escudero:2025}.
It was originally proposed to explicitly constrain $H_0$ without the sound horizon scale in \cite{Baxter:2020}, where the authors used the gravitational lensing of the CMB as the foundation of their measurement.
As discussed in \secref{theory}, CMB lensing is sensitive to the broad-band shape of the matter power spectrum (which depends on the matter-radiation equality scale, with comoving wavenumber \refedit{$\keq \propto \Om h^2$}).
The authors then tightened the resulting constraint by including additional $\Om$ information from uncalibrated\footnote{\refedit{In the sense of local-distance-ladder absolute distance calibration. See also \secref{theory_prev}.}} SNe Ia.
A recent constraint based on this method \cite{Farren:2024} combined CMB lensing data from ACT \cite{Das:2011,Sherwin:2017} and \Planck \cite{Planck:2014,Planck:2016lens,Planck:2020lens,Carron:2022} with galaxies from the unWISE catalog \cite{Schlafly:2019} and uncalibrated SNe Ia from Pantheon+ \cite{Brout:2022}, producing a precise sound horizon-independent measurement of $H_0 = 64.3^{+2.1}_{-2.4}\,\kmsMpc$.
\cwrminor{Likewise, in \cite{deBelsunce:2025}, DESI DR1 quasars were combined with ACT and \Planck CMB lensing data to obtain $H_0 = 69.1^{+2.2}_{-2.6}\,\kmsMpc$.}
Other approaches include determining $H_0$ directly from $\keq$ via the turnover scale of the matter power spectrum \cite{Bahr-Kalus:2023, Bahr-Kalus:2025}, or through the zero-crossing scale of the matter correlation function \cite{Prada:2011}. 
In \cite{Krolewski:2024, Krolewski:2025} it was demonstrated that $H_0$ can also be inferred from measurements of the cosmological energy densities without using sound-horizon information.
As mentioned previously, the method used in this work and in our previous analysis builds in turn on the techniques developed in \cite{Philcox:2021, Farren:2022, Philcox:2022}; in \cite{Philcox:2022}, the authors combined galaxy clustering from BOSS with \Planck CMB lensing and the Pantheon+ supernova sample, obtaining a 3.6\% constraint $H_0 = 64.8^{+2.2}_{-2.5} \kmsMpc$.

3-dimensional galaxy surveys, which cover large swathes of cosmic volume and probe many modes of the underlying cosmic density field, are one of the most promising avenues to constrain $H_0$ without the sound horizon.
The precision of this measurement will continue to increase as current and future surveys push to larger volumes; these include the Dark Energy Spectroscopic Instrument (DESI; \cite{DESI2016a.Science, DESI2016b.Instr,DESI2023a.KP1.SV}), Euclid \cite{Euclid:2024}, the Nancy Grace Roman Space Telescope \cite{Wang:2022}, and SPHEREx \cite{Dore:2015}. 

In this work, however, we demonstrate a significant step forward using data that is already available.
In our previous analysis, there were still opportunities to substantially improve the power of our measurement; we now incorporate two major additions.
The first of these is the consistent inclusion of density field-reconstructed BAO measurements from DESI DR1.
As we discuss in \secref{theory}, this ``uncalibrated'' BAO data primarily constrains $\Om$ and does not depend on the absolute size of the sound horizon scale, only assuming that it functions as a standard ruler\footnote{In the language of \cite{Brieden:2022}, constraints from these measurements are both ``uncalibrated'' (agnostic to the absolute size of the sound horizon) and ``unnormalized'' (rely on the signal's redshift evolution rather than amplitude).}.
Other sound horizon-independent $H_0$ measurements (e.g. \cite{Brieden:2023, Bahr-Kalus:2025}) have shown large improvements from the inclusion of similar information, making it an appealing addition to our analysis.
We also include a measurement of the CMB angular acoustic scale $\tstar = \rs/D_\mathrm{A}(z_*)$ as an additional uncalibrated BAO datapoint at \refedit{high redshift}, providing a long lever arm on the redshift evolution of the sound horizon feature.

Secondly, a key achievement of this work is its demonstrably robust constraints, in particular with respect to the inclusion of type-Ia supernovae.
In \cite{Zaborowski:2025}, our final result was heavily influenced by all three of the SN Ia datasets we considered, with their relatively high $\Om$ posteriors significantly shifting the joint $H_0$ constraint to lower values.
In that work we considered three SN Ia analyses: DES Y5 SN \cite{DES:2024sne}, Pantheon+ \cite{Brout:2022}, and Union3 \cite{Rubin:2023}, which give $\Om$ constraints $\Om = 0.352 \pm 0.017$, $0.334 \pm 0.018$, and $0.356^{+0.028}_{-0.026}$, respectively.
While at face value this \refedit{strong dependence on SNe Ia} is not a problem, it is important to confirm these results using alternative, independent datasets.
\cwrminor{Another reason to consider analysis variants independent of supernovae is that the main DESI DR1 cosmology analyses \cite{DESI2024.VI.KP7A, DESI2024.VII.KP7B} found that the combination DESI + CMB + SNe Ia is best fit by an evolving dark energy model, with a shifted $\Om$ \refeditt{(and lower $H_0$)} preference relative to the \lcdm best fit, \refeditt{warranting comparison with other probes}.}
Therefore, to test the robustness of our results, we showcase combinations with two new datasets: a recent cross-correlation analysis \cite{Farren:2024} of unWISE galaxies with \planckact CMB lensing, and the Dark Energy Survey's Year 3 ``6$\times$2pt'' joint analysis with CMB lensing from the South Pole Telescope and \Planck (see \secref{theory_ext}).
As we will discuss, these datasets are both inherently free of sound horizon information and serve as useful cross-checks against results that include supernovae.
For these robustness tests, as well as to facilitate comparison with our previous constraints, we include one SN Ia dataset in this work: the Dark Energy Survey Year 5 supernova analysis (DES Y5 SN; \cite{DES:2024sne}).
\cwrminor{This is the supernova dataset that in our previous work led to the largest shift in the $H_0$ posterior.}
We will show that with the inclusion of uncalibrated BAO/$\tstar$ measurements, our final results are highly consistent in all cases, hinting that the data are substantially more constraining than in our previous analysis.

This work represents a significant improvement in harnessing the full power of DESI DR1 to  constrain $H_0$ without the sound horizon, and is another step in laying the groundwork for high-precision measurements in future galaxy surveys.
After incorporating the improvements listed above, we achieve a sub-2\% sound horizon-independent measurement of $H_0$, made more exciting by its robustness across several external dataset combinations.
The remainder of this paper is laid out as follows.
In \secref{theory} we describe the theoretical concepts underpinning this work, and in \secref{data} we detail each specific dataset used.
In \secref{methodology} we give the various settings used in our likelihood and analysis pipeline, followed by results in \secref{results}.
Finally, we discuss our results in broader context in \secref{discussion} before concluding in \secref{conclusions}.

%%%%%%%%%%%%%%%%%%%%%%%%%%%%%%%%%%%%%%%%%%%%%%%%%%%%%%%%%%%%
\section{Theory and Methodology}
\label{sec:theory}
%%%%%%%%%%%%%%%%%%%%%%%%%%%%%%%%%%%%%%%%%%%%%%%%%%%%%%%%%%%%

In \secref{theory_prev} we review the core theory of our technique, which is shared with our previous work \cite{Zaborowski:2025}. In \secref{theory_bao} and \secref{theory_ext}, we explain in detail the novel aspects of the current work.

%%%%%%%%%%%%%%%%%%%%%%%%%%%%%%%%%%%%%%%%%%%%%%%%%%%%%%%%%%%%
\subsection{Core Theory}
\label{sec:theory_prev}

In the following, we largely follow the methodology of \cite{Zaborowski:2025}, and also direct the reader to Section 2 of that work.
The galaxy power spectrum contains two major features: the turnover at wavenumber $\keq \sim 0.015 \hmpcinv$ and its characteristic oscillations (or ``wiggles'') beginning at slightly larger $k$.
These features are sourced by the physics of the epoch of matter-radiation equality and recombination, respectively.
When calibrated, i.e. when the absolute physical sizes of these features are assumed to be known, these scales function as standard rulers.
That is, by observing each feature at different redshifts/cosmic epochs, one can infer the expansion history of the Universe.
This fact makes these two scales the most powerful sources of $H_0$ information in current galaxy surveys, with the sound horizon the better constrained of the two (the matter-radiation equality scale is more difficult to measure both due to cosmic variance as well as large-scale systematics; see \secref{datavec} and \secref{covariance} for some discussion of systematics mitigation in our pipeline).

In \lcdm the wavenumber corresponding to the horizon size at matter-radiation equality scale can be expressed as:
\begin{align}
    \keq \approx \left(2 \Obc H_0^2 z_{\mathrm{eq}}\right)^{1/2} \approx 7.46 \times 10^{-2} \Obc h \Theta_{2.7}^{-2} \,\left[\hmpcinv\right]
	\label{eq:keq}
\end{align}
\cite{Eisenstein:1998} \refedit{(note the factor of $h$ in the units)}, where $\Theta_{2.7} \equiv T_{\mathrm{CMB}} / (2.7\,\mathrm{K})$ and $\Obc$ is the sum of the baryon and cold dark matter density parameters $\Ob + \Ocdm$.
On the other hand, the sound horizon at the redshift of baryon-photon decoupling $z_\mathrm{d}$ (also known as the baryon drag epoch) is given by:
\begin{align}
    \nonumber
    \rd &\equiv \int_{z_d}^{\infty} \frac{\cs(z)}{H(z)} \,dz \\
    &\approx 147.05 h \left(\frac{\ob}{0.02236}\right)^{-0.13} \left(\frac{\obc}{0.1432}\right)^{-0.23} \left(\frac{\Neff}{3.04}\right)^{-0.1} \, \left[\hinvmpc\right]
	\label{eq:rs}
\end{align}
\cite{Brieden:2023}, where $\cs$ is the speed of sound in the photon-baryon plasma prior to decoupling.
In this expression we use physical densities $\omega_\mathrm{X} \equiv \Omega_\mathrm{X} h^2$.
\cwrminor{As can be seen from these relations, $H_0$ can be constrained from the observed matter-radiation equality feature\footnote{\refedit{Schematically, the observed matter-radiation equality feature is $\keq$ multiplied by the angular-diameter distance or Hubble distance at the tracer's redshift, which are in turn also mainly functions of $H_0$ and $\Obc$.}} given information on $\Obc$ (noting that $\Theta_{2.7}$ is well-constrained by CMB experiments), and likewise $H_0$ can be constrained from the sound horizon feature given information on $\Obc$, $\ob$, and $\Neff$ (which encodes the properties of photons and neutrinos, and is well-constrained by the CMB and big bang nucleosynthesis).}

\begin{figure}
\begin{center}
	\includegraphics[width=0.9\columnwidth]{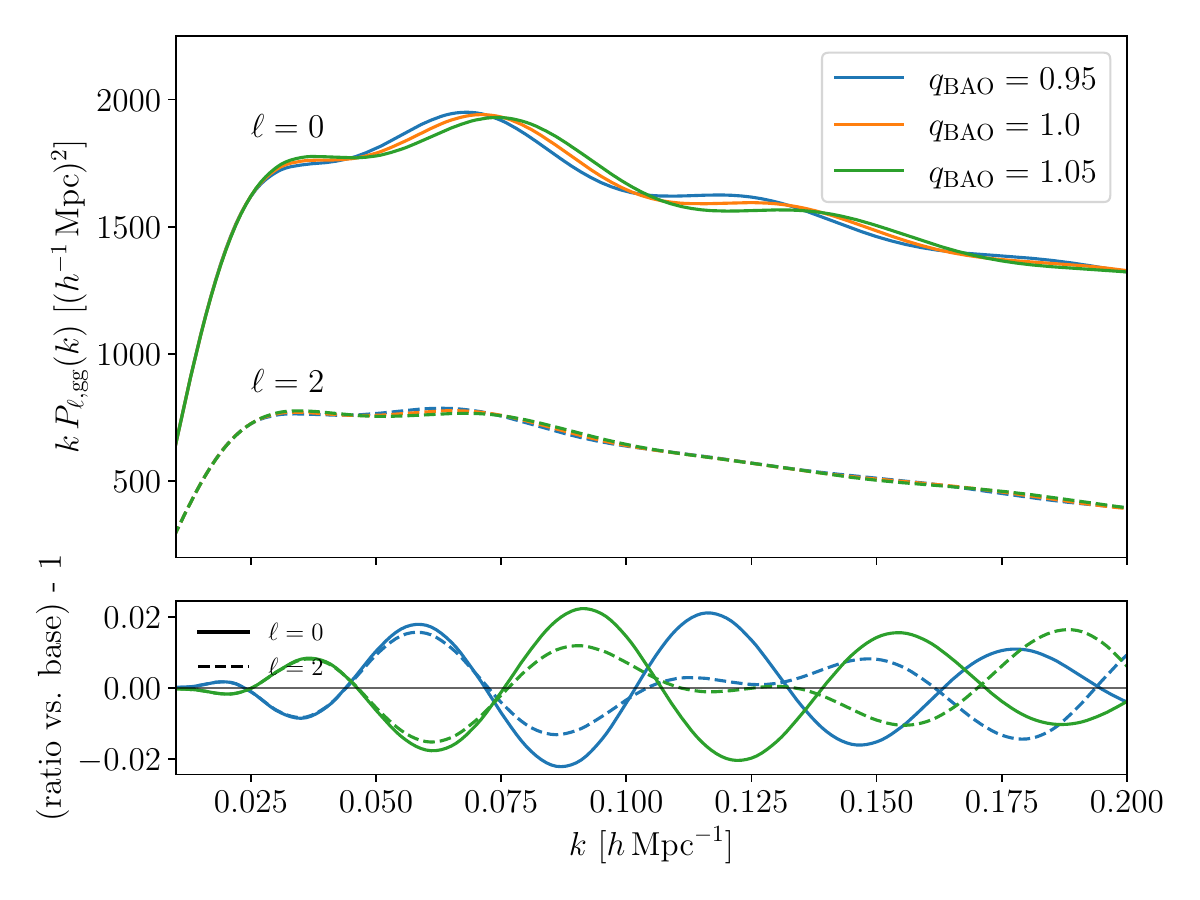}
    \caption{\textbf{Top:} The monopole ($\ell=0$; solid lines) and quadrupole ($\ell=2$; dashed lines) of the theoretical non-linear galaxy power spectrum are plotted for luminous red galaxies (LRGs) in the redshift range $0.4 < z < 0.6$.
    The wiggle-rescaling procedure detailed in \secref{theory_prev} is illustrated for several values of $\qbao$; $\qbao=0.95$ is plotted in blue, $\qbao=1.0$ in orange, and $\qbao=1.05$ in green.
    $\qbao=1$ corresponds to \lcdm, while larger values shift the BAO wiggles forward, and smaller values shift them backward, corresponding to a smaller/larger implied sound horizon scale respectively.
    \textbf{Bottom:} Residuals are shown versus the case $\qbao = 1$ for both multipoles $\ell = 0, 2$.
    }
    \label{fig:rescaling}
\end{center}
\end{figure}

Because an analysis of the full galaxy power spectrum naturally includes $H_0$ information from the sound horizon scale, we must remove this information from our measurement.
We accomplish this using a heuristic rescaling technique, originally developed in \cite{Farren:2022} and illustrated in \figref{rescaling}.
The idea of this technique is to first decompose the linear matter power spectrum into smooth and wiggly components, where the power spectrum wiggles can then be modified independently to tune the implied sound horizon scale.
To do this we use a simplified implementation\footnote{\url{https://github.com/cosmodesi/cosmoprimo/blob/25d486ef9a351c74b11fbd4eaafbc3cbceb21e2b/cosmoprimo/bao_filter.py\#L512}} of the method described in Appendix D of \cite{Brieden:2022}, which we very briefly summarize here.
The smooth component is obtained by first dividing the spectrum by the smooth Eisenstein–Hu formula \cite{Eisenstein:1998} to isolate an approximate oscillatory part.
The gradient of this residual is computed numerically, and its extrema, corresponding to BAO nodes, are identified.
The smooth curve is required to pass through these nodes, with two quadratic splines constructed separately for the maxima and minima.
Averaging these and re-multiplying by the Eisenstein–Hu broadband spectrum yields the final smooth spectrum, and the wiggly part is obtained by subtracting it from the original.
\cwrminor{The authors of \cite{Brieden:2022} explain that this method is more robust and accurate than only using the smooth Eisenstein–Hu formula, and has improved flexibility to handle model variations and extensions beyond \lcdm.}
\refedit{We have checked that any differences between this de-wiggling procedure and that used in \cite{Farren:2022,Philcox:2022} are highly subdominant to the DESI DR1 measurement uncertainties.}

After performing this separation procedure, we rescale only the wiggly component of the power spectrum by a free parameter $\qbao$ and recombine it with the smooth part:
\begin{align}
    P_{\mathrm{lin}}\left(k,\qbao\right) = P_{\mathrm{lin}}^{\mathrm{smooth}}\left(k\right) + P_{\mathrm{lin}}^{\mathrm{wiggly}}\left(\qbao\cwr{^{-1}} \cdot k\right)
\end{align}
This modified linear spectrum is then used as input to the non-linear galaxy clustering model described in \secref{theory_code}. 
This transformation rescales the effective sound horizon scale by a factor $\qbao^{-1}$\,, phenomenologically capturing any possible alterations in the physics that sets this scale. 
\figref{rescaling} illustrates the effect for a few representative values of $\qbao$;
$\qbao = 1$ corresponds to \lcdm, while larger (smaller) values shift the BAO features to higher (lower) $k$.
Allowing $\qbao$ to vary freely during inference introduces additional degeneracy between the sound horizon scale and the underlying cosmological parameters, ensuring that $H_0$ cannot be constrained from this information.
Previous tests \cite{Farren:2022,Philcox:2022} confirm that this procedure yields constraints independent of the sound horizon at the precision of DESI and Euclid.

As in \cite{Zaborowski:2025} we include additional sound horizon-independent information from big bang nucleosynthesis (BBN) and the gravitational lensing of the CMB.
In particular, in \cite{Baxter:2020} it was pointed out that CMB lensing is insensitive to the sound horizon scale because the large width of the lensing kernel causes this information to be washed out in projection along the line of sight.
Rather, CMB lensing is sensitive to the broadband shape of the matter power spectrum, thereby sourcing information from the matter-radiation equality scale.

It is apparent that we can break the matter-radiation equality degeneracy \refedit{\\$\keq \propto \Obc h^2 \sim \Om h^2$} by providing more information on $\Om$. 
In \cite{Zaborowski:2025} this was done using both i) the Alcock-Paczy\'{n}ski (AP) effect \cite{Alcock:1979} measured in the DESI 3-d Lyman-$\alpha$ (\lya) forest, which traces the structure of neutral hydrogen through its absorption of \lya photons emitted by quasars, as well as ii) uncalibrated type-Ia supernovae, which trace the expansion history of the Universe via their luminosity distance vs. redshift relationship.
Conceptually, both of these constrain $\Om$ because they are measures of \textit{relative} distances.
Cosmological distances typically involve integrals over the Hubble parameter:
\begin{align}
    H(z) = H_0 \sqrt{\Om (1+z)^3 + \Or (1+z)^4 + \OL}~,
\label{eq:Hz}
\end{align}
where $\Or$ is the cosmological density parameter for relativistic species (well-constrained from the CMB temperature), and $\OL = 1 - \Om - \Or$ (in flat \lcdm) corresponds to the energy density due to the cosmological constant.
\refedit{For simplicity, we have neglected massive neutrinos in this expression;} it can be seen in any case that the $H_0$ factor drops out in relative distances, leaving a residual dependence on $\Om$.

To be more specific, the AP effect arises when converting from redshift-space coordinates (RA, Dec, $z$) to Cartesian coordinates ($x_1$, $x_2$, $x_3$).
To perform this transformation, a fiducial cosmological model must be assumed, and any discrepancy with the true cosmology of the Universe will cause an apparent distortion along the line of sight.
That is, a feature that is spherical in reality will instead appear stretched or squashed along the line of sight.
The AP effect then constrains the true $\Om$ as the value that restores relative isotropy.
SNe Ia, on the other hand, measure distances via their observed fluxes.
SNe Ia are thought to be standardizable candles, with similar intrinsic brightnesses arising from common physical mechanisms.
If the cosmic distance ladder (e.g. \cite{Riess:2022,Freedman:2024}) is used to calibrate their shared absolute luminosity, then each supernova's observed flux implies an \textit{absolute} physical distance, and $H_0$ can be constrained.
Left uncalibrated, their fluxes still imply 
a \textit{relative} distance relation, the shape of which as a function of redshift is sensitive to $\Om$.

In this work we again use the DESI \lya AP effect to better constrain $\Om$,\footnote{AP information is also automatically included in the clustering of DESI galaxies. However, the signal is stronger in the \lya forest because \refedit{i) the relevant ratio $F_{\mathrm{AP}} \equiv D_{\mathrm{M}}(z) / D_{\mathrm{H}}(z) = D_{\mathrm{M}}(z) H(z)/c$ is more sensitive to $\Om$ at high redshift, and ii) it has high density and a large range of usable modes, particularly extending to smaller scales \cite{Cuceu:2021}.}} and we include an analysis variant featuring the DES Y5 SN dataset (\secref{data_sn}) as both a point of comparison and robustness check.

%%%%%%%%%%%%%%%%%%%%%%%%%%%%%%%%%%%%%%%%%%%%%%%%%%%%%%%%%%%%
\subsection{New in This Work: Uncalibrated BAO and $\tstar$}
\label{sec:theory_bao}

The first major addition to this work is the inclusion of uncalibrated BAO data.
In a standard BAO analysis (e.g. \cite{DESI2024.VI.KP7A, DESI.DR2.BAO.cosmo}), the measured BAO feature is used to constrain two distance ratios, corresponding to its extent in angular (across the line of sight) and redshift space (along the line of sight).
The angular scale of the BAO can be expressed as the ratio $\Delta \theta = \rd / \DM(z)$, where in flat \lcdm $\DM(z)$ is the comoving angular diameter distance
\begin{align}
    \DM(z) = \int_{0}^{z} \frac{c \,dz^\prime}{H(z^\prime)}~.
\end{align}
On the other hand, the BAO feature along the line of sight can be expressed as $\Delta z = \rd / 
D_{\mathrm{H}}(z)$, where $D_{\mathrm{H}}(z)$ is the Hubble distance
\begin{align}
    D_{\mathrm{H}}(z) = \frac{c}{H(z)}~.
\end{align}
In cases where signal-to-noise is limited, these distances can be combined into an angle-averaged quantity:
\begin{align}
    \DV(z) \equiv \left( z \DM(z)^2 D_{\mathrm{H}}(z) \right)^{1/3}~.
\end{align}
These define the primary observables measurable from the BAO feature.
It is common to normalize the observables by their fiducial values under a certain cosmology; we perform this normalization using the DESI fiducial cosmology, i.e. the mean values of the \Planck 2018 \lcdm parameters\footnote{This is also referred to as the \texttt{c000} cosmology in the AbacusSummit simulations: \url{https://abacussummit.readthedocs.io/en/latest/cosmologies.html}} \cite{Planck:2020}.
Thus we perform our analysis using the normalized BAO dilation parameters (see also details in \secref{datavec}):
\begin{align}
    \alpha_\parallel &\equiv \frac{\DHrd}{\left(\DHrd\right)_{\mathrm{fid}}} \label{bao_dilation1}\\
    \alpha_\perp &\equiv \frac{\DMrd}{\left(\DMrd\right)_{\mathrm{fid}}} \label{bao_dilation2}\\
    \alpha_\mathrm{iso} &\equiv \frac{\DVrd}{\left(\DVrd\right)_{\mathrm{fid}}}~.
    \label{bao_dilation3}
\end{align}

Importantly, the BAO observables by themselves constrain only the parameter $\Om$ and the combination $H_0 \rd$.
In order to measure $H_0$ alone, the BAO standard ruler must first be ``calibrated'' so that the absolute size of $\rd$ is known.
Referring back to Equation \ref{eq:rs}, this can be done, for example, by enforcing a prior on $\ob$ from BBN.
In the absence of this calibration, constraints from BAO provide no information on $H_0$ (as it is totally degenerate with $\rd$), but still however provide information on $\Om$, making this an appealing addition to further break the $\Om - H_0$ degeneracy in our sound horizon-marginalized measurement.

\begin{figure}
\begin{center}
	\includegraphics[width=0.8\columnwidth]{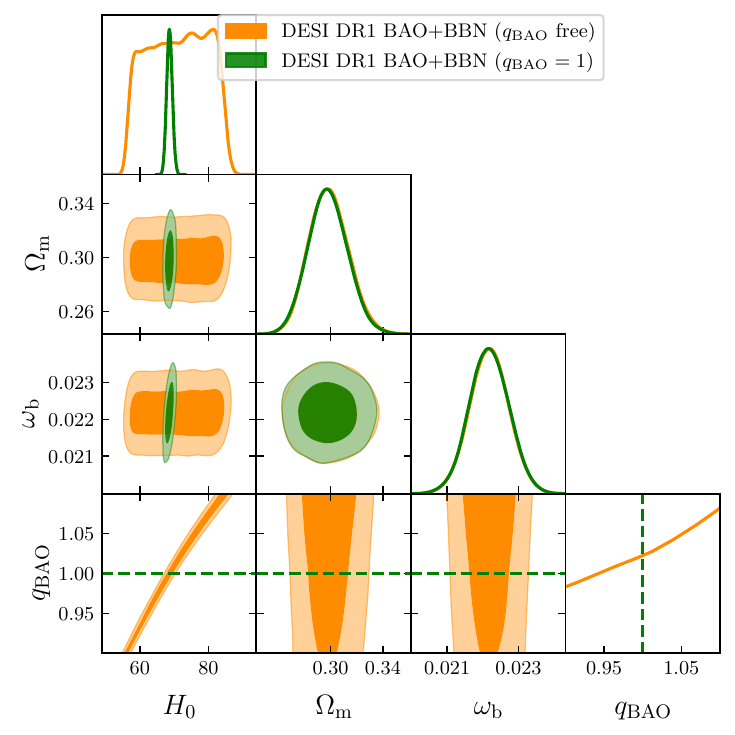}
    \caption{Constraints on the parameters $H_0$, $\Om$, $\ob$, and $\qbao$ using the DESI DR1 BAO measurements and a BBN prior on $\ob$.
    In green: contours obtained when fixing $\qbao = 1$.
    As expected, these constraints closely match those obtained in the main DESI DR1 BAO analysis \cite{DESI2024.VI.KP7A}.
    In orange: contours obtained when allowing the parameter $\qbao$ to vary.
    In the latter case, $H_0$ is no longer constrained, but importantly the $\Om$ constraints obtained in both cases are nearly identical.
    }
    \label{fig:uncal_bao}
\end{center}
\end{figure}

While we include an $\ob$ prior in this work, which would normally calibrate the sound horizon standard ruler and thus constrain $H_0$ via the absolute sound horizon scale, we account for this using the same rescaling parameter $\qbao$ described in the previous section.
\cwrminor{That is, when calculating theoretical values of the BAO dilation parameters (Equations \ref{bao_dilation1} - \ref{bao_dilation3}), we substitute $\rd \to \rd / \qbao$} \refedit{in the model under investigation (but not the fiducial cosmology)}.
In \figref{uncal_bao} we demonstrate this explicitly.
We show constraints from DESI DR1 BAO+BBN using our modified pipeline, both for fixed $\qbao = 1$ (green) and varied $\qbao$ (orange).
As expected, the green contours (with $\qbao$ fixed to 1) closely match the official DESI DR1 BAO+BBN constraints \cite{DESI2024.VI.KP7A}.
However, when $\qbao$ is allowed to vary, $H_0$ becomes unconstrained even though we have included a prior on $\ob$.
Importantly, the $\Omega_m$ constraints obtained in both cases match very well, as expected from the discussion above.
In order to include the DESI DR1 BAO data in our measurement, we properly account for the correlation between the BAO and full-shape data using the same covariance and methodology as in the main DESI full-shape+BAO analysis \cite{DESI2024.V.KP5, DESI2024.VII.KP7B}, described in more detail in \secref{covariance}.

We note that while it might be naively thought that the information of the DESI BAO measurements is fully contained in the DESI full-shape power spectra, there is in fact additional information in the BAO data, primarily due to density field reconstruction.\footnote{\cwrminor{We also note that because the DESI BAO datapoints are measured in configuration space, they are not generally constrained to contain information from the same range of Fourier modes as included in the power spectrum.}}
Reconstruction is a technique employed to sharpen the BAO feature by partially undoing the smearing caused by nonlinear bulk flows \cite{Eisenstein:2007, Padmanabhan:2012}.
In effect, this reconstruction reclaims some BAO information that has slipped into higher-point functions due to nonlinear structure formation.
The reconstruction algorithm used in the DESI DR1 BAO analysis \cite{DESI2024.VI.KP7A} is described in detail in \cite{KP4s4-Paillas}.

As mentioned previously, uncalibrated BAO measurements have been included in other sound horizon-free $H_0$ measurements, often providing significant improvements in precision \cite{Brieden:2022, Bahr-Kalus:2025}.
In the present analysis, we take this one step further, including a measurement of the CMB acoustic scale $\tstar$, which functions as an additional high-redshift angular BAO measurement.
We ensure this measurement remains ``uncalibrated'' by performing the sound horizon rescaling procedure analogously to the case of the DESI BAO measurements, with the caveat that the relevant theoretical quantity in calculating $\tstar$ is the sound horizon scale at recombination ($\rs$) rather than at baryon decoupling ($\rd$).
In practice, we assume that $\rs$ and $\rd$ both achieve their respective \lcdm values at $\qbao = 1$, and both rescale consistently according to the same factor $\qbao$.\footnote{\refedit{In most cosmologies, these scales vary similarly \cite{Pogosian:2020}.}}

%%%%%%%%%%%%%%%%%%%%%%%%%%%%%%%%%%%%%%%%%%%%%%%%%%%%%%%%%%%%
\subsection{New in This Work: External Datasets}
\label{sec:theory_ext}

As mentioned earlier, in this work we consider combinations with two supernova-independent datasets in order to evaluate the robustness of our final results.
Here we will describe relevant theoretical concepts from each analysis within the context of this work, whereas the more technical details of each dataset will be given in \secref{data}.

\refedit{
The first of these datasets is the Dark Energy Survey Year 3 (DES Y3) joint analysis with CMB lensing from the South Pole Telescope (SPT) and \Planck (hereafter referred to as the DES Y3 6$\times$2pt analysis \cite{Abbott:2023}; technical details in \secref{data_des6x2pt}).
The CMB lensing component of this analysis is automatically free of sound horizon information as discussed previously \cite{Baxter:2020}.
The remaining components which may in principle contain sound horizon information are the DES galaxy positions and weak-lensing shape distortions (or ``shear'').
In the case of cosmic shear (i.e. the shear auto-correlation), it has been shown that the BAO feature will be difficult to measure even for Stage IV photometric weak lensing surveys \cite{Ding:2019}.
This leaves primarily the galaxy clustering as a potential source of sound horizon information.
Perhaps the most straightforward check of this is simply to analyze the data using a prior on $\ob$, which as discussed previously would calibrate the sound horizon scale and thus constrain $H_0$ if there is a significant BAO signal. 
}

\refedit{
Using the publicly available\footnote{\url{https://des.ncsa.illinois.edu/releases/y3a2/Y3key-products}} DES Y3 chains, we reweight the samples according to our Gaussian $\ob$ and $\ns$ priors listed in Table \ref{table:priors}.
After doing this, we obtain the constraint $H_0 = 70.1^{+8.8}_{-8.4} \kmsMpc$ from the DES Y3 1$\times$2pt (galaxy clustering) dataset, compared to $68.0^{+3.9}_{-5.1} \kmsMpc$ for the 3$\times$2pt and $68.5^{+2.6}_{-2.3} \kmsMpc$ for the 6$\times$2pt results that we use in this work.
These results indicate that the $H_0$ constraint from the 6$\times$2pt dataset is not dominated by the sound horizon, as the 1$\times$2pt / galaxy clustering dataset would be the largest potential source of this information, and we have found that it does not significantly constrain $H_0$ on its own.
We note that in \cite{DES:2024bao} a $\sim$2\% constraint on the BAO scale was obtained using the DES Year 6 galaxy clustering.
However, they used a BAO-optimized sample for this work, including narrower redshift binning to reduce projection effects ($\Delta z = 0.1$), in addition to modified angular binning and scale cuts, to successfully resolve the feature.
}

\refedit{
The second new dataset considered in this work is a recent cross-correlation analysis \cite{Farren:2024} of \planckact CMB lensing with the projected clustering of unWISE galaxies (hereafter referred to as ``\actcrossunwise''; described further in \secref{data_actplanck}).
Each component of this analysis is sound horizon-free for similar reasons as given above; \refeditt{in particular, the redshift binning of the unWISE galaxies ($\Delta z \sim 1$) is much wider than in the \dessixcrosstwopt analysis, while the angular binning is comparable.}
Importantly, we note that the \dessixcrosstwopt and \actcrossunwise datasets are not independent, as both contain \Planck CMB lensing.
}

%%%%%%%%%%%%%%%%%%%%%%%%%%%%%%%%%%%%%%%%%%%%%%%%%%%%%%%%%%%%
\section{Data}
\label{sec:data}
%%%%%%%%%%%%%%%%%%%%%%%%%%%%%%%%%%%%%%%%%%%%%%%%%%%%%%%%%%%%

In this section we describe in detail the specific datasets used in this work.

\subsection{Galaxies and the Lyman-$\alpha$ Forest}
\label{sec:data_desi}

The Dark Energy Spectroscopic Instrument (DESI) is a multi-object spectrographic survey utilizing $\sim$5,000 robotically positioned fibers and operating on the Mayall 4-meter telescope at Kitt Peak National Observatory \cite{DESI2022.KP1.Instr,DESI2016b.Instr,FocalPlane.Silber.2023,Corrector.Miller.2023}.
\cwrminor{DESI is currently conducting an eight-year survey that will cover roughly 17,000 $\deg^2$ of the sky.
The full survey will lead to 63 million spectroscopically-confirmed galaxies and quasars, compared to the initial forecasts of 39 million \cite{DESI2016a.Science}, and it aims to probe the nature of dark energy via the most precise measurement of the expansion history of the Universe to date \cite{Snowmass2013.Levi}.}
The spectroscopic data reduction pipeline used for DESI is described in \cite{Spectro.Pipeline.Guy.2023}, while the survey operations and observation plan are detailed in \cite{SurveyOps.Schlafly.2023}.
DESI target selection is based on imaging data from the public DESI Legacy Imaging Surveys \cite{LS.Overview.Dey.2019}, and the DESI scientific program was validated in \cite{DESI2023a.KP1.SV}, accompanied by an early data release \cite{DESI2023b.KP1.EDR}.

The first-year data release of the DESI survey (DESI DR1 or DESI 2024; \cite{DESI2024.I.DR1}) contains the positions and spectra of over 5.7 million unique objects, which are classified among several tracers in multiple redshift bins.
Several science Key Papers accompanied DESI DR1: two-point clustering measurements and validation \cite{DESI2024.II.KP3}, measurements of the BAO feature in galaxies and quasars \cite{DESI2024.III.KP4} and the Lyman-$\alpha$ (\lya) forest \cite{DESI2024.IV.KP6}, and the full-shape study of galaxies and quasars \cite{DESI2024.V.KP5}.
Cosmological interpretations were given for the BAO measurements \cite{DESI2024.VI.KP7A} and the full-shape analysis \cite{DESI2024.VII.KP7B}.
Recently, measurements and constraints from BAO in DESI DR2 were given in \cite{DESI.DR2.BAO.lya, DESI.DR2.BAO.cosmo}, although we note that in the present work we exclusively use data from DESI DR1.

In this work we jointly analyze the full-shape power spectrum of DESI galaxies together with density field-reconstructed BAO measurements.
The full-shape power spectrum is measured for a total of six unique tracer/redshift combinations in DESI DR1: the low-redshift bright galaxy sample (BGS) at redshift range $0.1 < z < 0.4$, luminous red galaxies (LRGs) in three redshift bins ($0.4 < z < 0.6$, $0.6 < z < 0.8$, and $0.8 < z < 1.1$), emission line galaxies (ELGs) in the redshift range $1.1 < z < 1.6$, and quasars (quasi-stellar objects or QSOs) in the redshift range $0.8 < z < 2.1$, together comprising a total of 4.7 million galaxies.
The post-reconstruction BAO is also measured for each of these tracer/redshift pairs, as well as for the \lya forest in the redshift range $1.8 < z < 4.2$.
While we do not include the full-shape power spectrum of the \lya forest in our analysis, we do use the measurement by \cite{Cuceu:2025} of the AP effect observed in the latter.

\subsection{External Dataset 1: $\tstar$}
\label{sec:data_thetastar}

We use the measurement of the CMB acoustic scale $100 \tstar = 1.04110 \pm 0.00031$ reported in the \Planck 2018 analysis \cite{Planck:2020}, using the combined TT,TE,EE+lowE+lensing dataset and the \texttt{Plik} likelihood.\footnote{\refedit{During review, it was pointed out to us by the referee that when combining this $\tstar$ measurement with CMB lensing, we may be double-counting information. Given that the corresponding lensing-free constraint from TT,TE,EE+lowE is very similar ($100 \tstar = 1.04109 \pm 0.00030$), we do not expect a large impact from this in our results.}}

\subsection{External Dataset 2: DES Year 5 Supernova Analysis}
\label{sec:data_sn}

The Dark Energy Survey Year 5 supernova analysis (DES Y5 SN; \cite{DES:2024sne}) represents the largest and most homogeneous single sample of high-redshift supernovae to date; it combines novel observations of 1,635 photometrically classified SNe Ia (spanning $0.1 < z < 1.3$) with existing observations of 194 low-redshift ($0.025 < z < 0.1$) SNe Ia.
Because uncalibrated SNe Ia primarily constrain $\Om$ (as was noted in \secref{theory_ext}), we choose to extract a marginalized $\Om$ posterior to be used as a prior in our analysis. 

\subsection{External Dataset 3: \planckact Lensing ($\times$ unWISE galaxy Clustering)}
\label{sec:data_actplanck}

The \actcrossunwise dataset (described and analyzed in \cite{Farren:2024}) uses a combination of CMB lensing data from the \Planck PR4 lensing analysis (covering $\sim$27,600  $\deg^2$) \cite{Carron:2022}, and the ACT DR6 lensing reconstruction (9,400 $\deg^2$) \cite{Madhavacheril:2024,Qu:2024}.
The unWISE catalog \cite{Krolewski:2020, Schlafly:2019} is distilled from observations by the full-sky Wide-field Infrared Survey Explorer (WISE; 
\cite{Wright:2010}), and also includes imaging from the NEOWISE survey extension \cite{Mainzer:2011} and post-hibernation NEOWISE-Reactivation (NEOWISER) mission \cite{Mainzer:2014}.
\cwrminor{The \actcrossunwise analysis uses a 3$\times$2pt dataset, incorporating cross-correlations between CMB lensing and the projected clustering of unWISE galaxies, in addition to both auto-correlations.}
\cwrminor{In this work we include the constraints obtained in \cite{Farren:2024} at the posterior level, as discussed in \secref{importance}.}

Of course, the \actcrossunwise dataset is a superset of the \planckact CMB lensing measurements, which were also employed in our previous analysis.
Therefore, in this work we additionally present results using only the \planckact lensing data for comparison purposes.
For this, we use the publicly available\footnote{\url{https://github.com/ACTCollaboration/act\_dr6\_lenslike}} \planckact CMB-marginalized lensing-only likelihood, enforcing the same settings as our previous work: \\\texttt{variant = actplanck\_baseline} and \texttt{lens\_only = true}\,.

\subsection{External Dataset 4: DES 6x2pt Dataset}
\label{sec:data_des6x2pt}

The \dessixcrosstwopt analysis, presented in \cite{Abbott:2023}, combines photometry from the Year 3 data release of the Dark Energy Survey (DES Y3; \cite{Flaugher:2005, Abbott:20223x2}) with CMB lensing measurements from the South Pole Telescope (SPT; \cite{Carlstrom:2011}) and \Planck.
The joint CMB lensing map constructed from the 2,500 $\deg^2$ SPT-SZ field \cite{Story:2013} and \Planck 2015 data release \cite{Planck:2016dr} is described in \cite{Omori:2017, Omori:2023}.

The basic quantities measured from these component datasets are the DES galaxy positions (giving the overdensity 
$\delta_\mathrm{g}$) and shapes (giving the lensing $\gamma$), and the SPT+\Planck CMB lensing convergence $\kappa_\mathrm{CMB}$.
As evident in its name, the \dessixcrosstwopt analysis constructs and analyzes six 2-point functions from these quantities.
These are: galaxy clustering $\left<\delta_\mathrm{g} \delta_\mathrm{g}\right>$, galaxy--galaxy lensing $\left<\delta_\mathrm{g} \gamma\right>$, \refedit{shear--shear correlation $\left<\gamma \gamma\right>$,}  galaxy density $\times$ CMB lensing $\left<\delta_\mathrm{g} \kappa_\mathrm{CMB}\right>$, galaxy shear $\times$ CMB lensing $\left<\gamma \kappa_\mathrm{CMB}\right>$, and finally the CMB lensing auto-correlation $\left<\kappa_\mathrm{CMB} \kappa_\mathrm{CMB}\right>$.
\cwrminor{We include the constraints obtained in \cite{Abbott:2023} at the posterior level, as discussed in \secref{importance}.}

%%%%%%%%%%%%%%%%%%%%%%%%%%%%%%%%%%%%%%%%%%%%%%%%%%%%%%%%%%%%
\section{Likelihood and Analysis}
\label{sec:methodology}
%%%%%%%%%%%%%%%%%%%%%%%%%%%%%%%%%%%%%%%%%%%%%%%%%%%%%%%%%%%%

Here we describe the various components of our analysis pipeline.
\refedit{In the following we assume a cosmology with fixed $\tau_{\mathrm{reio}}=0.0544$ (motivated by \emph{Planck} 2018 \cite{Planck:2020}) when modeling DESI measurements, which are insensitive to changes in this parameter.
As in \cite{Planck:2020}, we also assume three degenerate species of massive neutrinos and the standard-model prediction $N_{\mathrm{eff}}=3.046$.}

%%%%%%%%%%%%%%%%%%%%%%%%%%%%%%%%%%%%%%%%%%%%%%%%%%%%%%%%%%%%
\subsection{Data Vector}
\label{sec:datavec}

For the full-shape part of the data vector, we use the DESI DR1 galaxy power spectrum multipoles for each tracer (except the \lya forest) over $k \in \left[0.02, 0.20\right] \hmpcinv$ for $\ell = 0, 2$, in steps of $\Delta k = 0.005\hmpcinv$.
This aligns with the analysis choices of the main DESI analyses \cite{DESI2024.V.KP5, DESI2024.VII.KP7B}, as well as our previous work \cite{Zaborowski:2025}.
Data processing follows \cite{DESI2024.II.KP3}, including: (i) truncation of small angular scales \cite{Pinon:2024}; (ii) “rotation” of the data vector to compactify the window in $k$-space \cite{Pinon:2024}; (iii) corrections for the radial (RIC) \cite{DeMattia:2019} and angular (AIC) \cite{DESI2024.II.KP3} integral constraints.
The RIC arises from matching the redshift distribution of randoms to the data; the AIC from imaging systematic weights obtained by regressing galaxy density on imaging-property maps. 
Corrections for the RIC and AIC are derived from differences between mean power spectra of mocks with/without the effect, fit with a polynomial.

For the post-reconstruction BAO part of the data vector, we use measurements of the normalized BAO dilation parameters $\alpha_\parallel$ and $\alpha_\perp$ for each DESI DR1 tracer, except for the BGS and QSO samples, where only the isotropic parameter $\alpha_\mathrm{iso}$ is used due to low signal-to-noise.
The measurements of these parameters are described in detail in \cite{DESI2024.III.KP4}.

%%%%%%%%%%%%%%%%%%%%%%%%%%%%%%%%%%%%%%%%%%%%%%%%%%%%%%%%%%%%
\subsection{Covariance Matrix}
\label{sec:covariance}

We model the covariance matrix as the sum of statistical and systematic components, following the methodology of \cite{DESI2024.VII.KP7B}.
The power spectrum statistical term \cite{KP4s6-Forero-Sanchez}, as well as the cross-covariance between the power spectrum and post-reconstruction BAO measurements, is estimated from 1,000 EZmocks \cite{Chuang:2015}.
These mocks are designed to match DESI DR1 clustering statistics while being faster to generate than full N-body simulations. 
\cwrminor{We additionally apply the multiplicative correction factors described in \cite{DESI2024.V.KP5}, which account for differences in the mock-based and analytical covariance estimates.}
The BAO-only part of the covariance matrix comes from fits to the data for the BAO dilation parameters \cite{DESI2024.VII.KP7B}. 
The systematic term in the full-shape part of the covariance \cite{DESI2024.V.KP5} includes any effect producing a parameter bias $\ge 0.2\sigma$ (in DESI DR1 error units), such as imaging systematics, fiber assignment, and HOD-dependent effects.
The combination procedure is described in Appendix D of \cite{DESI2024.V.KP5}.
\cwrminor{Adding these components in quadrature yields a total systematic uncertainty of $\sim 0.46\sigma$ relative to the DESI DR1 error, well below the sound-horizon–marginalized error level used in this work.
The systematic part of the post-reconstruction BAO covariance, which includes theoretical
modeling uncertainties, uncertainties due to the galaxy-halo connection, and observational systematic effects is detailed in \cite{DESI2024.III.KP4}.
Added in quadrature, the systematic uncertainties are conservatively estimated to be 0.245\% and 0.3\% for the isotropic and anisotropic BAO dilation parameters, respectively.}

%%%%%%%%%%%%%%%%%%%%%%%%%%%%%%%%%%%%%%%%%%%%%%%%%%%%%%%%%%%%
\subsection{Theory Model}
\label{sec:theory_code}

For the full-shape part of our theory, we compute the theoretical galaxy power spectrum using the publicly available \texttt{velocileptors} code\footnote{\url{https://github.com/sfschen/velocileptors}} \cite{Chen:2020,Chen:2021}, which implements the effective field theory of large-scale structure (EFT of LSS).
The Lagrangian perturbation theory (LPT) model is used to third order, with the third-order bias $b_3$ fixed to zero following \cite{DESI2024.V.KP5,DESI2024.VII.KP7B}.
We adopt the reparameterization of EFT biases, counterterms, and stochastic terms described in \cite{DESI2024.V.KP5}, which allows physically interpretable priors to be applied to these parameters (\secref{priors}).
The BAO part of our theory utilizes the Boltzmann code \texttt{CLASS} \cite{Blas:2011} to compute the relevant theoretical quantities.

%%%%%%%%%%%%%%%%%%%%%%%%%%%%%%%%%%%%%%%%%%%%%%%%%%%%%%%%%%%%
\subsection{Priors}
\label{sec:priors}

\begin{table}
\centering
\begin{tabular}{c c c}
 \hline
 Parameter & Prior & Notes \\
 \hline
 $H_0$ & $\mathrm{Unif}\left[20, 100\right]$ & [$\kmsMpc$] \\
 $\Om$ & $\mathrm{Unif}\left[0.01, 1.0\right]$ & \\ 
 $\ob$ & $\mathcal{N}\left(0.02218, 0.00055^2\right)$ & BBN \cite{Schoeneberg:2024} \\ 
 $\logA$ & $\mathrm{Unif}\left[1.61, 3.91\right]$ & \\ 
 $\ns$ & $\mathcal{N}\left(0.9649, 0.042^2\right)$ & $10\times$ \Planck \cite{Planck:2020} \\
 $\qbao$ & $\mathrm{Unif}\left[0.9, 1.1\right]$ & \\ 
 \hline
 $\left(1+b_1\right) \sigma_8$ & $\mathrm{Unif}\left[0.0, 3.0\right]$ & \\
 $b_2 \sigma_8^2$, $b_s \sigma_8^2$ & $\mathcal{N}\left(0, 5^2\right)$ & \\
 $\alpha_0$, 
 $\alpha_2$ & $\mathcal{N}\left(0, 12.5^2\right)$ & +Approx. Jeffreys prior\tablefootnote{\label{jeffreys_note}Jeffreys prior \cite{Jeffreys:1946} applied in addition to the listed Gaussian prior. In practice, we use an approximation to the Jeffreys prior demonstrated in \cite{Hadzhiyska:2023}.} \\
 $\mathrm{SN}_0$ & $\mathcal{N}\left(0, 2^2\right) \times 1/\bar{n}_{\mathrm{g}}$ & +Approx. Jeffreys prior$^{\ref{jeffreys_note}}$ \\
 $\mathrm{SN}_2$ & $\mathcal{N}\left(0, 5^2\right) \times f_{\mathrm{sat}} \sigma_{1\,\mathrm{eff}}^2/\bar{n}_{\mathrm{g}}$ & +Approx. Jeffreys prior$^{\ref{jeffreys_note}}$ \\ 
 \hline
\end{tabular}
\caption{
Priors are shown for constraints obtained from the DESI data alone.
We adopt the physically-motivated nuisance parameterization detailed in \cite{DESI2024.V.KP5} and \cite{KP5s2-Maus}.
Per \cite{KP5s2-Maus}, we multiply the prior width of parameter $\mathrm{SN}_0$ by the Poissonian shot noise $1/\bar{n}_{\mathrm{g}}$, and for parameter $\mathrm{SN}_2$ we multiply the prior width by the Poissonian shot noise times the expected satellite galaxy fraction times the mean satellite velocity dispersion, corresponding to the characteristic velocity dispersion.
}
\label{table:priors}
\end{table}

We adopt priors largely following \cite{DESI2024.V.KP5,DESI2024.VII.KP7B}, with differences noted below.
The baseline Bayesian priors for DESI-only constraints are listed in \tabref{priors}.
We sample the cosmological parameters $H_0$, $\Om$, $\ob$, $\logA$, $\ns$, and $\qbao$, as well as bias parameters $(1+b_1)\sigma_8$, $b_2\sigma_8^2$, and $b_s\sigma_8^2$ (with this bias parameterization chosen to reduce projection effects \cite{DESI2024.V.KP5}).
Counterterms $\alpha_{[0,2]}$ and stochastic terms $\mathrm{SN}_{[0,2]}$ are treated with a maximization procedure approximating a Jeffreys prior \cite{Jeffreys:1946,Hadzhiyska:2023} (see discussion in \cite{Zaborowski:2025} and below).
We note that we have fixed the higher-order parameters $\alpha_{4}$ and $\mathrm{SN}_{4}$ to zero, in alignment with the main DESI analyses \cite{DESI2024.V.KP5, DESI2024.VII.KP7B}.
These parameters were freed in our previous work \cite{Zaborowski:2025}, leading to negligibly wider constraints.
Differences with the main analyses \cite{DESI2024.V.KP5,DESI2024.VII.KP7B} still include sampling $\Om$ rather than $\ocdm$, and the partial Jeffreys prior applied to the nuisance parameters.
As in our previous work, we fix the sum of neutrino masses to $\sumnu = 0.06\,\mathrm{eV}$.

We include two external priors: a BBN prior on $\ob$ \cite{Schoeneberg:2024}, and a wide prior on $\ns$ ($\mathcal{N}(0.9649,0.042^2)$) which is centered on the \Planck 2018 TT,TE,EE+lowE+lensing constraint but with $10\times$ the width of the posterior \cite{Planck:2020}.  
When combining with CMB lensing, we instead adopt $\ns \sim \mathcal{N}(0.96,0.02^2)$, in accordance with the \Planck lensing analysis \cite{Planck:2020lens}.
Gaussian priors are applied to $\alpha_{[0,2]}$ so that the $1\sigma$ EFT correction at $k_{\max}=0.2\hmpcinv$ is $\le 50\%$ of the total signal, corresponding to $\mathcal{N}(0,12.5^2)$ \cite{DESI2024.V.KP5}.
For stochastic terms, we use $\mathcal{N}(0,2^2)\times1/\bar{n}$ for $\mathrm{SN}_0$ (twice the Poisson shot noise), and $\mathcal{N}(0,5^2)\times f_{\mathrm{sat}}\sigma_{1,\mathrm{eff}}^2/\bar{n}$ for $\mathrm{SN}_2$.

The approximate Jeffreys prior is applied to counterterms and stochastic terms to mitigate projection effects \cite{Simon:2023,Holm:2023} in the high-dimensional EFT nuisance space, which can bias cosmological parameters, especially when extending beyond \lcdm (e.g. when freeing $\qbao$).
Without this prior, mock DESI full-shape analyses in \cite{Zaborowski:2025} showed $H_0$ shifts of $\sim 2\sigma$, which are largely removed once the Jeffreys prior is enforced.
We implement this prior following the procedure outlined in \cite{Hadzhiyska:2023}, fixing nuisance parameters to their maximum-posterior values at each point in cosmological+bias parameter space.
The authors of that work showed that this technique is approximately equivalent to marginalizing over these parameters in the presence of a Jeffreys prior.

%%%%%%%%%%%%%%%%%%%%%%%%%%%%%%%%%%%%%%%%%%%%%%%%%%%%%%%%%%%%
\subsection{Emulator}
\label{sec:emulator}

Direct evaluation of the full EFT model requires several seconds per call, making it impractical for a single Markov-chain Monte Carlo run (MCMC; \cite{Robert:2004}) and prohibitive for repeated testing.
To accelerate the computation, we use the \texttt{desilike} 4th-order Taylor-series emulator\footnote{\url{https://desilike.readthedocs.io/en/latest/api/emulators.html\#module-desilike.emulators.taylor}} to generate approximate power spectrum multipoles for each tracer. 
\texttt{desilike} isolates the cosmology-dependent part of the perturbation theory integrals -- responsible for most of the computational cost -- from the nuisance-dependent part, allowing the Taylor expansion to be performed only over the cosmological parameters \refedit{(including $\qbao$)}.
We also include the BAO part of the theory when training our emulators in order to ensure consistency during evaluation, \refedit{such that the modeled BAO observables are also 4th-order Taylor expansions in the cosmological parameters}.

%%%%%%%%%%%%%%%%%%%%%%%%%%%%%%%%%%%%%%%%%%%%%%%%%%%%%%%%%%%%
\subsection{Sampling}
\label{sec:sampling}

We perform MCMC inference with the \texttt{desilike} implementation of the No-U-Turn Sampler (NUTS) \cite{Hoffman:2014} on the Perlmutter cluster at the National Energy Research Scientific Computing Center (NERSC).
Sampling proceeds until the maximum-eigenvalue Gelman-Rubin statistic \cite{Brooks:1998} satisfies $|R - 1| < 0.03$, and an effective sample size of 10,000, computed as the total chain length divided by the integrated correlation length. 
Chains are generated using the emulated power spectrum + post-reconstruction BAO model, and the true posterior is recovered via importance sampling, reweighting each chain element by the ratio of the true to emulated posterior probabilities.\footnote{When re-evaluating with the true model, we use the same maximization procedure for counterterms and stochastic terms described in \secref{priors}.}

%%%%%%%%%%%%%%%%%%%%%%%%%%%%%%%%%%%%%%%%%%%%%%%%%%%%%%%%%%%%
\subsection{Combination with External Datasets}
\label{sec:importance}

We use various techniques to combine each external dataset with the baseline DESI power spectrum + post-reconstruction BAO analysis.
For CMB lensing, we combine at the likelihood level by reweighting our MCMC chains with the \planckact likelihood (\secref{data_actplanck}).
We note that this is permissible because the covariance between the projected statistics and the DESI multipoles is negligible \cite{Taylor:2022rgy}.
For the DES Y5 SN dataset, the $\Om$ constraints are Gaussian, enabling an analytic reweighting by the posterior of their analysis.
In contrast, the DESI \lya AP constraints on $\Om$ are non-Gaussian; in this case we use \texttt{CombineHarvesterFlow}\footnote{\url{https://github.com/pltaylor16/CombineHarvesterFlow}} \cite{Taylor:2024eqc} to obtain the joint constraints at the posterior level.
\cwrminor{\texttt{CombineHarvesterFlow} uses normalizing flows (reviewed in \cite{Kobyzev:2021}) to estimate a reweighting from two independent MCMC chains that gives their joint posterior distribution.}
We also use this tool to perform combinations with the \actcrossunwise and \dessixcrosstwopt datasets, which we note are uncorrelated with the DESI measurements.
Finally, constraints including $\tstar$ are jointly sampled along with the DESI data so that the $\qbao$ rescaling can be performed consistently.

%%%%%%%%%%%%%%%%%%%%%%%%%%%%%%%%%%%%%%%%%%%%%%%%%%%%%%%%%%%%
\section{Results}
\label{sec:results}
%%%%%%%%%%%%%%%%%%%%%%%%%%%%%%%%%%%%%%%%%%%%%%%%%%%%%%%%%%%%

\begin{table}
\centering
\begin{tabular}{c c c}
 \hline
 Dataset(s) & $H_0$ [$\kmsMpc$] & $\qbao$ \\
 \hline
 DESI FS & $72.1^{+4.5}_{-4.4}$ & $1.016^{+0.038}_{-0.037}$ \\
 DESI FS + $\Om^{\mathrm{Ly}\alpha\mathrm{AP}}$ & $70.5^{+3.3}_{-3.4}$ & $1.003^{+0.027}_{-0.028}$ \\
 All DESI (FS + $\Om^{\mathrm{Ly}\alpha\mathrm{AP}}$ + Uncal. BAO) & $70.2 \pm 2.6$ & $1.011 \pm 0.019$ \\
 All DESI + $\tstar$ & $70.8^{+2.0}_{-2.2}$ & $1.018^{+0.012}_{-0.013}$ \\
 \hline
 All DESI + $\tstar$ + \planckact Lensing & $70.5^{+1.6}_{-1.8}$ & $1.016^{+0.009}_{-0.010}$ \\
 All DESI + $\tstar$ + \actcrossunwise & $\mathbf{69.2^{+1.3}_{-1.4}}$ & $\mathbf{1.009 \pm 0.009}$ \\
 All DESI + $\tstar$ + \dessixcrosstwopt & $\mathbf{70.3^{+1.4}_{-1.2}}$ & $\mathbf{1.016 \pm 0.008}$ \\
 \hline
 All DESI + $\tstar$ + \planckact Lensing + DES Y5 SN & \cwr{$\mathbf{69.6^{+1.3}_{-1.8}}$} & \cwr{$\mathbf{1.014^{+0.007}_{-0.010}}$} \\
 All DESI + $\tstar$ + \actcrossunwise + DES Y5 SN & \cwr{$68.7^{+1.1}_{-1.3}$} & \cwr{$1.008 \pm 0.007$} \\
 All DESI + $\tstar$ + \dessixcrosstwopt + DES Y5 SN & \cwr{$69.8 \pm 1.3$} & \cwr{$1.015^{+0.009}_{-0.008}$} \\
 % ... & ...\\
 \hline
\end{tabular}
\caption{Constraints on the parameters $H_0$ and $\qbao$ for various dataset combinations considered in this work.
Shown in bold are the ``headline'' constraints of this analysis, which are discussed in \secref{discussion}.
All listed results are 68\% credible intervals and include a BBN prior.
}
\label{table:results}
\end{table}

\begin{figure}
\begin{center}
	\includegraphics[width=0.8\columnwidth]{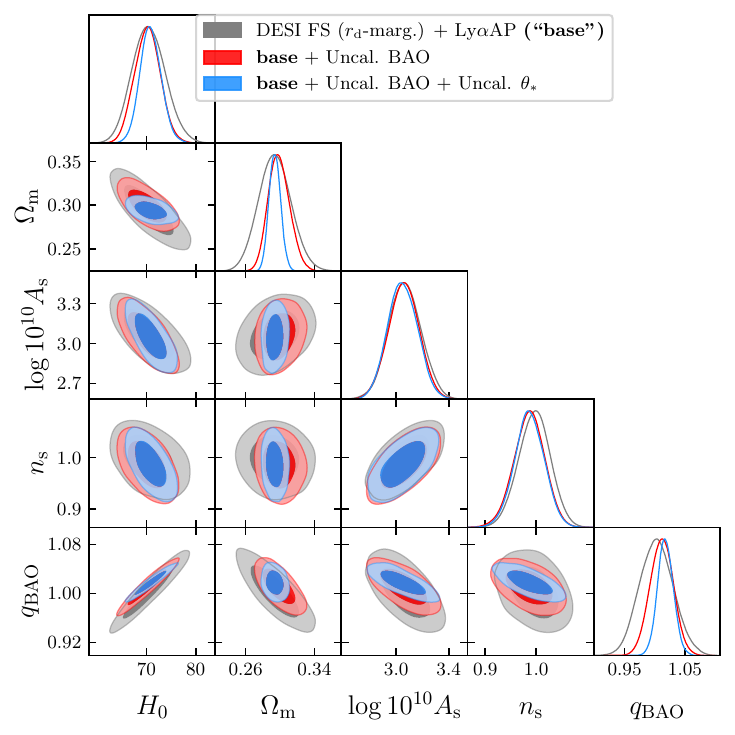}
    \caption{
    Contours are shown for the parameters $H_0$, $\Om$, $\logA$, $\ns$, and $\qbao$.
    In gray: constraints from the DESI DR1 full-shape galaxy clustering and the \lya AP effect.
    This dataset also formed the base of our previous analysis \cite{Zaborowski:2025} and serves as a point of comparison.
    In red: same as gray, but additionally including DESI DR1 uncalibrated post-reconstruction BAO measurements.
    In blue: same as red, but also including a measurement of the CMB acoustic scale $\tstar$ from \Planck 2018 \cite{Planck:2020}.
    All results shown include a BBN prior.
    }
    \label{fig:desi_fs_bao_thetastar}
\end{center}
\end{figure}

\begin{figure}
\begin{center}
	\includegraphics[width=0.8\columnwidth]{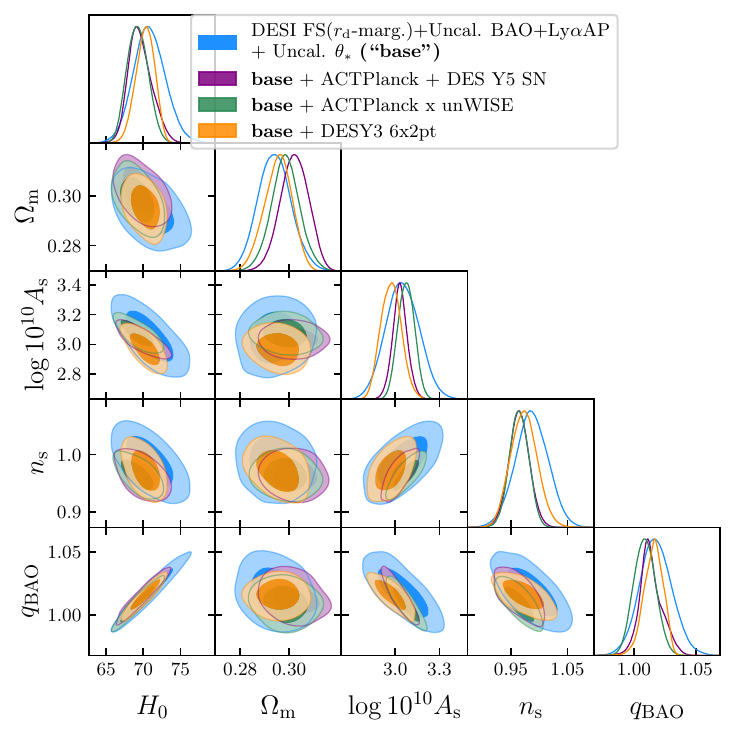}
    \caption{
    Contours are shown for the parameters $H_0$, $\Om$, $\logA$, $\ns$, and $\qbao$.
    \refedit{Note that the contour labeled ``base'' is now different from that in \figref{desi_fs_bao_thetastar}.}
    In blue: constraints are shown for the full DESI DR1 dataset (full-shape galaxy clustering + uncalibrated post-reconstruction BAO + the \lya AP effect), plus the uncalibrated CMB acoustic scale $\tstar$ measured in \Planck 2018 \cite{Planck:2020}.
    We note that this is the same as the blue contour in the previous \figref{desi_fs_bao_thetastar}.
    In purple: same as blue, but additionally including \planckact CMB lensing and the DES Y5 SN dataset.
    This dataset combination (excepting uncalibrated BAO and $\tstar$) was also used in our previous work \cite{Zaborowski:2025} and serves as a point of comparison.
    In green: same as blue, but also including the \actcrossunwise \cwrminor{3$\times$2pt} dataset (\secref{data_actplanck}).
    In orange: same as blue, but also including the \dessixcrosstwopt dataset (\secref{data_des6x2pt}).
    All results shown include a BBN prior.
    }
    \label{fig:full_results}
\end{center}
\end{figure}

We tabulate our various sound horizon-marginalized $H_0$ and $\qbao$ constraints in \tabref{results}, and in \figref{desi_fs_bao_thetastar} and \figref{full_results} we show 2D contours for the parameters $\left\{H_0, \Om, \logA, \ns, \qbao \right\}$.

We begin in \figref{desi_fs_bao_thetastar} by illustrating the major improvements in this work relative to \cite{Zaborowski:2025}, namely the inclusion of DESI uncalibrated post-reconstruction BAO and $\tstar$ measurements.
In gray are contours from the DESI full-shape measurements along with \lya AP information, which served as the base of our previous work and give a point of comparison.\footnote{We note that this constraint is actually $\sim 4$\% broader than in our previous analysis \cite{Zaborowski:2025} due to a now-fixed numerical issue in the BGS emulator which led to inaccurate predictions for large deviations from $\qbao=1$ (differences in the final constraints after combining all datasets were negligible).}
Layered on top of these full-shape constraints, we add contours that include the uncalibrated post-reconstruction DESI BAO (red) and uncalibrated $\tstar$ (blue) measurements.
It can be seen that the addition of uncalibrated BAO and $\tstar$ provide in turn approximately equal improvements to the precision of the resulting $H_0$ constraint, going from $\sim 4.8$\% precision from the power spectrum + \lya AP measurements to $\sim 3.7$\% when including uncalibrated BAO, and finally to a precision of $\sim 3.0$\% when including uncalibrated $\tstar$.
Excitingly, the precision achieved by this latter dataset (DESI FS ($\rd$ marg.) + \lya AP + Uncal. BAO + Uncal. $\tstar$) already rivals the best results from our previous work, before considering any additional external datasets.
Further, this constraint is largely internal to DESI, with relatively minimal external assumptions from BBN and $\tstar$, whereas our previous constraints included both CMB lensing and supernovae.
Interestingly, we see that this current constraint, while roughly as precise as the previous one, prefers somewhat higher $H_0$ than in our previous work.
We also note a slight (1-2$\sigma$) preference for $\qbao > 1$, which corresponds to a diminished sound horizon scale relative to the base \lcdm assumption.

In \figref{full_results}, we show our headline constraints from this work, which include the full combinations with each of the external datasets discussed in \secref{data}.
For reference, the blue contour in this plot (DESI DR1 + $\tstar$) is the same as in the previous \figref{desi_fs_bao_thetastar}.
Layered on top of this base dataset, we show three alternative data combinations: \planckact CMB lensing + DES Y5 SN (purple), \planckact CMB lensing x unWISE projected clustering (green), and \dessixcrosstwopt (orange).
Again, we note that these are not mutually independent (the latter two both contain \Planck CMB lensing).
There are several points that we would like to draw attention to in this plot.
Firstly, the combination with DES Y5 SN (the purple contours) was included as a point of comparison with our previous work \cite{Zaborowski:2025}; in that work, our $\rd$-marginalized $H_0$ constraint using DESI FS + \lya AP + \planckact lensing + DES Y5 SN was $66.7^{+1.7}_{-1.9}\,\kmsMpc$.
In this work, our corresponding constraint is \cwr{$69.6^{+1.3}_{-1.8}\,\kmsMpc$}, with the primary difference being the inclusion of uncalibrated post-reconstruction BAO and $\tstar$.
Interestingly, the constraints in this work seem to be more robust to the inclusion of supernovae, as the contours do not shift quite as severely toward lower values of $H_0$.
Additionally, we note the high level of consistency displayed between all three of the alternative final constraints.
Together, we interpret this improvement in robustness and consistency as an indication that this updated dataset is more powerful and highly constraining, pulling the resulting posteriors toward the same final result.

We emphasize that each of our final $H_0$ constraints achieves roughly 2\% precision or better; this is the state-of-the-art regime for sound horizon-free $H_0$ measurements from LSS.

%%%%%%%%%%%%%%%%%%%%%%%%%%%%%%%%%%%%%%%%%%%%%%%%%%%%%%%%%%%%
\section{Discussion}
\label{sec:discussion}
%%%%%%%%%%%%%%%%%%%%%%%%%%%%%%%%%%%%%%%%%%%%%%%%%%%%%%%%%%%%

\begin{figure}
\begin{center}
	\includegraphics[width=\columnwidth]{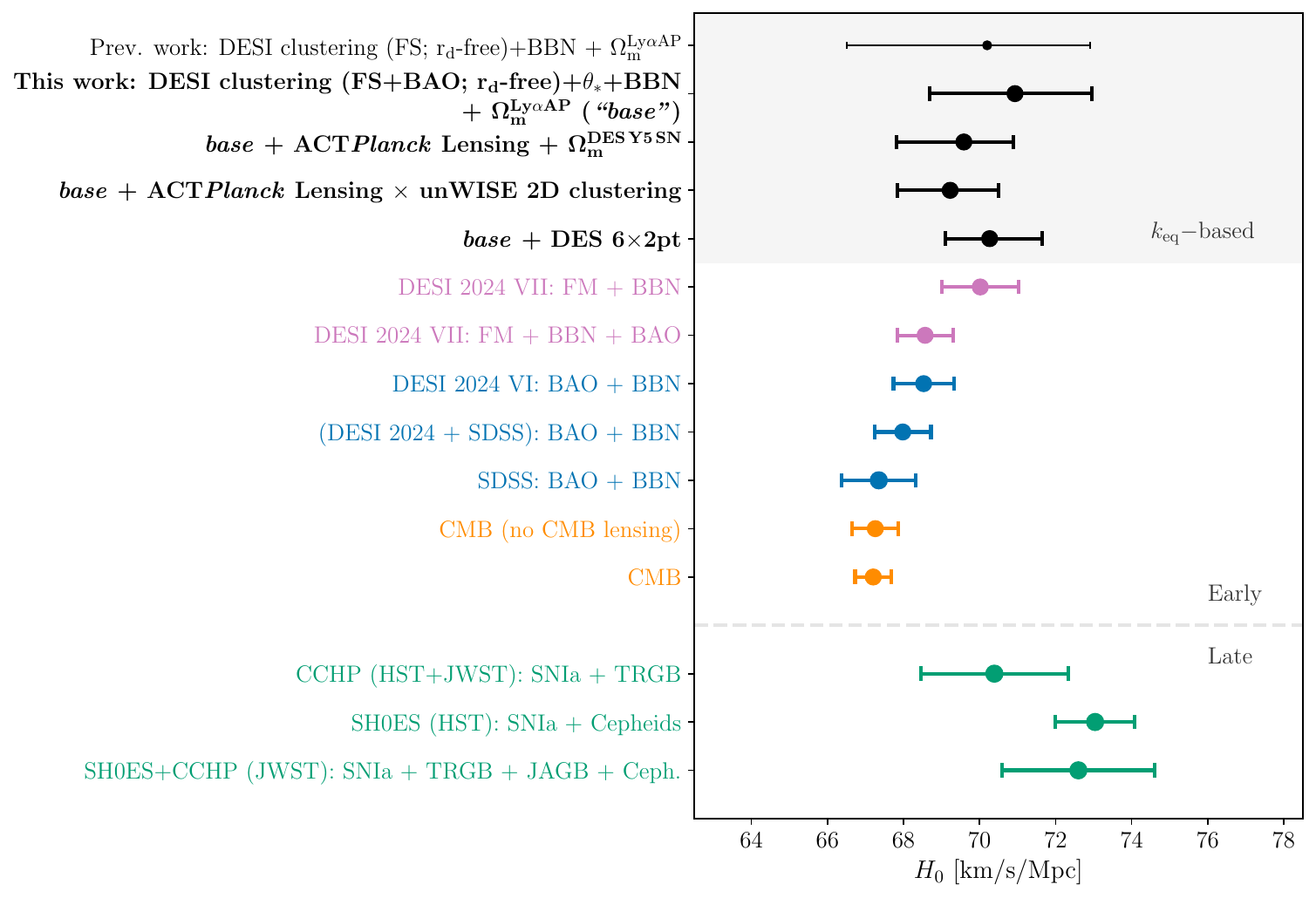}
    \caption{
    A whisker plot is shown of 68\% confidence interval constraints for several representative measurements of the Hubble constant.
    From top to bottom:
    In black: $\keq$-based constraints from this work and our previous analysis \cite{Zaborowski:2025}.
    In pink: constraints are shown from the main DESI DR1 full-shape + BAO analysis \cite{DESI2024.V.KP5}.
    In blue: constraints from the main DESI DR1 BAO-only analysis \cite{DESI2024.VI.KP7A}.
    In orange: $H_0$ constraints from the combined \planckact CMB dataset.
    In green: late-time $H_0$ constraints from: i) the Chicago-Carnegie Hubble Program (CCHP; \cite{Freedman:2024}), who used data from the Hubble Space Telescope (HST) and the James Webb Space Telescope (JWST), ii) the SH0ES team \cite{Riess:2022}, who used data from HST, as well as iii) combined constraints from the largest combination of both subsamples available in JWST \cite{Riess:2024}.
    JAGB stands for the J-region Asymptotic Giant Branch calibration method, and TRGB stands for calibration using the Tip of the Red Giant Branch  (see references for additional details).
    }
    \label{fig:whisker}
\end{center}
\end{figure}

We present an updated landscape of $H_0$ measurements in \figref{whisker} (compare to Figure 5 of \cite{Zaborowski:2025}).
At the top, shown in black are sound horizon-marginalized constraints from both this work and our previous analysis \cite{Zaborowski:2025}.
As these measurements largely rely on the broad-band shape of the matter power spectrum, the matter-radiation scale $\keq$ is the primary (but not only \cite{Jun-Qian:2025}) source of $H_0$ information.
The first, non-bolded whisker (DESI DR1 $\rd$-marginalized full-shape + \lya AP effect) was obtained in the previous work and is shown for comparison.
Directly following this is the corresponding constraint from this work, which includes DESI DR1 uncalibrated post-reconstruction BAO measurements, as well as an uncalibrated $\tstar$ measurement from \Planck 2018 \cite{Planck:2020}.
The final three constraints in black are further combined with the following respective datasets: i) \planckact lensing + DES Y5 SN, ii) \actcrossunwise, and iii) \dessixcrosstwopt.

As remarked previously, an important difference relative to our previous analysis is that when including uncalibrated BAO and $\tstar$ measurements, our constraints tend toward higher values of $H_0$, even when including SNe Ia.
We also now observe a higher level of consistency between the constraints from DESI data alone and those that include full combinations with external datasets.
We interpret this shift and overall consistency as being due to significantly more constraining data.

The updated results in \figref{whisker} lead to an interesting new conclusion.
Previously, our $\keq$-based $H_0$ constraints were highly consistent with the lower values preferred by the CMB, showing a residual tension with SH0ES at up to $\sim 3\sigma$.
On the other hand, our new constraints now lie near the middle of the two, and we note that our constraint including the DES Y5 SN dataset is actually statistically closer to the SH0ES constraint than it is to the \planckact CMB measurement.
We can no longer say with confidence that our data show that the Hubble tension exists independently of the sound horizon scale.

We also note another, more speculative trend in the plot: it appears that the $H_0$ measurements that are most reliant on the \lcdm prediction of the sound horizon scale (i.e. the CMB and BAO-only constraints) prefer the lowest values of $H_0$ and show the largest deviation with late-time SNe Ia constraints, while on the other hand our measurements which rely mostly on broad-band information prefer higher $H_0$ values.
Interestingly, a seemingly continuous trend of intermediate measurements connects the two, roughly aligning with their relative reliance on sound horizon vs. broad-band $H_0$ information.

Combined with our slight ($1-2\sigma$) observed preference for $\qbao > 1$, corresponding to a diminished sound horizon scale (a prediction of some Hubble tension-ameliorating models such as early dark energy / EDE), we should carefully consider this trend and see if it may have a physical explanation.
In \cite{Farren:2022}, the authors performed a similar $\rd$-marginalized analysis to this work using Euclid-like simulations, both for \lcdm and EDE underlying cosmologies.
They found that performing the analysis assuming \lcdm (as we have) when the true underlying cosmology is EDE tends to push sound horizon-based $H_0$ constraints towards higher values relative to their sound horizon-marginalized counterparts. 
Interestingly, this is directionally different from the tentative trend that we observe in \figref{whisker}, and is apparently at odds with our observed preference for a diminished sound horizon scale.
We plan to further investigate these results in the context of specific early-Universe physics models in a follow-up work; at a 2\% level of precision, we are entering a regime where any differences between $\keq$-based and $\rs$-based $H_0$ measurements will need to be explained, whether systematic or physical.

Looking ahead, the methodology presented here lays the foundation for future improvements.
Upcoming DESI data releases, covering a larger cosmic volume, will lead to substantially improved constraints simply by probing more large-scale modes.
However, there are also other avenues to further improve the constraints independently of survey volume.
A promising possibility is better pinning down uncertainties on the nuisance terms in the analysis, such as the EFT bias parameters and counterterms.
This could be achieved, for example, by including information from the cross-correlations between DESI galaxies and CMB lensing.
Another possibility is incorporating physically motivated priors on the nuisance terms, such as those based on halo occupation distribution (HOD) models \cite{Ivanov:2024,Zhang:2025}.
We plan to investigate these improvements in future work.

%%%%%%%%%%%%%%%%%%%%%%%%%%%%%%%%%%%%%%%%%%%%%%%%%%%%%%%%%%%%
\section{Conclusions}
\label{sec:conclusions}
%%%%%%%%%%%%%%%%%%%%%%%%%%%%%%%%%%%%%%%%%%%%%%%%%%%%%%%%%%%%

In this work, we have constrained the Hubble constant $H_0$ \refeditt{in the \lcdm cosmological model} independently of the sound horizon size at recombination.
At the core of our measurement (\secref{theory_prev}) is a wiggle-rescaling procedure applied to the theoretical galaxy power spectrum that effectively marginalizes over the absolute size of the sound horizon.
By incorporating uncalibrated post-reconstruction BAO measurements and the CMB acoustic scale $\tstar$, we have significantly tightened constraints relative to our previous analysis \cite{Zaborowski:2025}.

Our full results were shown in \secref{results}.
Combining measurements from DESI DR1 \cite{DESI2024.I.DR1} with minimal external information from BBN \cite{Schoeneberg:2024} and $\tstar$ \cite{Planck:2020}, we obtained $H_0 = 70.8^{+2.0}_{-2.2} \kmsMpc$, a $\sim$3\% constraint that already rivals our previous work.
We then importantly demonstrated the robustness of our full analysis by considering several external dataset combinations.
Combining the above with \planckact CMB lensing $\times$ unWISE galaxy clustering \cite{Farren:2024} gives $H_0 = 69.2^{+1.3}_{-1.4} \kmsMpc$; combination with the \dessixcrosstwopt analysis \cite{Abbott:2023} gives $H_0 = 70.3^{+1.4}_{-1.2} \kmsMpc$; including \planckact CMB lensing + the DES Y5 SN analysis \cite{DES:2024sne} gives \cwr{$69.6^{+1.3}_{-1.8}\,\kmsMpc$}.
These results are highly consistent and achieve better than 2\% precision on average, a major milestone for sound horizon-free cosmology.

Several key conclusions emerge, which were discussed in \secref{discussion}.
Relative to our previous work, we observe a shift to slightly higher values of $H_0$, reducing both the apparent tension with local distance-ladder measurements as well as the consistency with the CMB.
This weakens our prior conclusion that there is no evidence of new physics in DESI DR1, which might be expected to impact the sound horizon scale differently than the broad-band power spectrum / matter-radiation equality scale.
In fact, we observe a mild preference (1-2$\sigma$) for $\qbao>1$, corresponding to a diminished sound horizon scale (a feature predicted by some Hubble tension-ameliorating models), though we note that further investigation is required.
We expect a dedicated work in the future exploring these results in the context of specific models of early-Universe physics\refeditt{, and potentially relaxing \lcdm modeling assumptions (e.g. allowing for late-time dynamical dark energy)}.

The future of this measurement is bright.
Forthcoming data from surveys like DESI and Euclid will allow increasingly precise sound horizon-independent $H_0$ measurements, and in \secref{discussion} we highlighted cross-correlations and physically-informed priors as opportunities to further refine the analysis.
Ultimately, these advances will allow us to rigorously test whether the Hubble tension reflects new physics or residual systematics, providing an exacting probe of the standard cosmological model.

%%%%%%%%%%%%%%%%%%%%%%%%%%%%%%%%%%%%%%%%%%%%%%%%%%%%%%%%%%%%
\section{Data Availability}
%%%%%%%%%%%%%%%%%%%%%%%%%%%%%%%%%%%%%%%%%%%%%%%%%%%%%%%%%%%%

DESI Data Release 1 is available at: \url{https://data.desi.lbl.gov/doc/releases/dr1/}.

%%%%%%%%%%%%%%%%% ACKNOWLEDGEMENTS %%%%%%%%%%%%%%%%%%%%%
% \section*{Acknowledgements}
\acknowledgments

We thank Eoin \'{O} Colg\'{a}in, Ravi Sharma, Julien Lesgourgues, and Recai Erdem for helpful comments on the pre-print version of this manuscript.

This material is based upon work supported by the U.S. Department of Energy (DOE), Office of Science, Office of High-Energy Physics, under Contract No. DE–AC02–05CH11231, and by the National Energy Research Scientific Computing Center, a DOE Office of Science User Facility under the same contract. Additional support for DESI was provided by the U.S. National Science Foundation (NSF), Division of Astronomical Sciences under Contract No. AST-0950945 to the NSF’s National Optical-Infrared Astronomy Research Laboratory; the Science and Technology Facilities Council of the United Kingdom; the Gordon and Betty Moore Foundation; the Heising-Simons Foundation; the French Alternative Energies and Atomic Energy Commission (CEA); the National Council of Humanities, Science and Technology of Mexico (CONAHCYT); the Ministry of Science, Innovation and Universities of Spain (MICIU/AEI/10.13039/501100011033), and by the DESI Member Institutions: \url{https://www.desi.lbl.gov/collaborating-institutions}. Any opinions, findings, and conclusions or recommendations expressed in this material are those of the author(s) and do not necessarily reflect the views of the U. S. National Science Foundation, the U. S. Department of Energy, or any of the listed funding agencies.

This material is based upon work supported by the U.S. Department of Energy, Office of Science, Office of Workforce Development
for Teachers and Scientists, Office of Science Graduate Student Research (SCGSR) program.
The SCGSR program is administered by the Oak Ridge Institute for Science and Education (ORISE) for the DOE. ORISE is managed by ORAU under contract number DESC0014664.
All opinions expressed in this paper are the author’s and do not necessarily reflect the policies and views of DOE,
ORAU, or ORISE.

The DESI Legacy Imaging Surveys consist of three individual and complementary projects: the Dark Energy Camera Legacy Survey (DECaLS), the Beijing-Arizona Sky Survey (BASS), and the Mayall z-band Legacy Survey (MzLS). DECaLS, BASS and MzLS together include data obtained, respectively, at the Blanco telescope, Cerro Tololo Inter-American Observatory, NSF’s NOIRLab; the Bok telescope, Steward Observatory, University of Arizona; and the Mayall telescope, Kitt Peak National Observatory, NOIRLab. NOIRLab is operated by the Association of Universities for Research in Astronomy (AURA) under a cooperative agreement with the National Science Foundation. Pipeline processing and analyses of the data were supported by NOIRLab and the Lawrence Berkeley National Laboratory. Legacy Surveys also uses data products from the Near-Earth Object Wide-field Infrared Survey Explorer (NEOWISE), a project of the Jet Propulsion Laboratory/California Institute of Technology, funded by the National Aeronautics and Space Administration. Legacy Surveys was supported by: the Director, Office of Science, Office of High Energy Physics of the U.S.\ Department of Energy; the National Energy Research Scientific Computing Center, a DOE Office of Science User Facility; the U.S.\ National Science Foundation, Division of Astronomical Sciences; the National Astronomical Observatories of China, the Chinese Academy of Sciences and the Chinese National Natural Science Foundation. LBNL is managed by the Regents of the University of California under contract to the U.S.\ Department of Energy. The complete acknowledgments can be found at \url{https://www.legacysurvey.org/}.

Any opinions, findings, and conclusions or recommendations expressed in this material are those of the author(s) and do not necessarily reflect the views of the U.S.\ National Science Foundation, the U.S.\ Department of Energy, or any of the listed funding agencies.

The authors are honored to be permitted to conduct scientific research on I'oligam Du'ag (Kitt Peak), a mountain with particular significance to the Tohono O’odham Nation.

%%%%%%%%%%%%%%%%%%%%%%%%%%%%%%%%%%%%%%%%%%%%%%%%%%

%%%%%%%%%%%%%%%%%%%% REFERENCES %%%%%%%%%%%%%%%%%%

\bibliographystyle{JHEP}
\bibliography{biblio.bib, biblio_old, biblio_desi_20251012}

%%%%%%%%%%%%%%%%% APPENDICES %%%%%%%%%%%%%%%%%%%%%

% \appendix
% \section{Additional materials}

% Appendix...

\end{document}

%% file: DESI-2025-0540_authors_2025.10.21.tex
\author[1,2,3]{{E.~A.~Zaborowski}\orcidlink{0000-0002-6779-4277},}
\author[3]{{P.~Taylor},}
\author[1,2,3]{{K.~Honscheid}\orcidlink{0000-0002-6550-2023},}
\author[4]{{A.~Cuceu}\orcidlink{0000-0002-2169-0595},}
\author[5]{{A.~de~Mattia}\orcidlink{0000-0003-0920-2947},}
\author[6,7,8]{{A.~Krolewski},}
\author[1,2,3]{{M.~Rashkovetskyi}\orcidlink{0000-0001-7144-2349},}
\author[1,9,3]{{A.~J.~Ross}\orcidlink{0000-0002-7522-9083},}
\author[3]{{C.~To},}
\author[4]{{J.~Aguilar},}
\author[10]{{S.~Ahlen}\orcidlink{0000-0001-6098-7247},}
\author[4]{{A.~Anand}\orcidlink{0000-0003-2923-1585},}
\author[11]{{S.~BenZvi}\orcidlink{0000-0001-5537-4710},}
\author[12,13]{{D.~Bianchi}\orcidlink{0000-0001-9712-0006},}
\author[14]{{D.~Brooks},}
\author[15,16]{{F.~J.~Castander}\orcidlink{0000-0001-7316-4573},}
\author[4]{{T.~Claybaugh},}
\author[17]{{A.~de la Macorra}\orcidlink{0000-0002-1769-1640},}
\author[18,19]{{J.~Della~Costa}\orcidlink{0000-0003-0928-2000},}
\author[14]{{P.~Doel},}
\author[4,20]{{S.~Ferraro}\orcidlink{0000-0003-4992-7854},}
\author[21]{{A.~Font-Ribera}\orcidlink{0000-0002-3033-7312},}
\author[22,23]{{J.~E.~Forero-Romero}\orcidlink{0000-0002-2890-3725},}
\author[15,24,16]{{E.~Gaztañaga}\orcidlink{0000-0001-9632-0815},}
\author[25]{{G.~Gutierrez},}
\author[26,5]{{H.~K.~Herrera-Alcantar}\orcidlink{0000-0002-9136-9609},}
\author[27]{{C.~Howlett}\orcidlink{0000-0002-1081-9410},}
\author[28,29]{{D.~Huterer}\orcidlink{0000-0001-6558-0112},}
\author[30]{{M.~Ishak}\orcidlink{0000-0002-6024-466X},}
\author[19]{{R.~Joyce}\orcidlink{0000-0003-0201-5241},}
\author[31]{{D.~Kirkby}\orcidlink{0000-0002-8828-5463},}
\author[4]{{T.~Kisner}\orcidlink{0000-0003-3510-7134},}
\author[4]{{A.~Kremin}\orcidlink{0000-0001-6356-7424},}
\author[14]{{O.~Lahav},}
\author[3]{{C.~Lamman}\orcidlink{0000-0002-6731-9329},}
\author[4]{{M.~Landriau}\orcidlink{0000-0003-1838-8528},}
\author[32]{{L.~Le~Guillou}\orcidlink{0000-0001-7178-8868},}
\author[33,21]{{M.~Manera}\orcidlink{0000-0003-4962-8934},}
\author[1,9,3]{{P.~Martini}\orcidlink{0000-0002-4279-4182},}
\author[19]{{A.~Meisner}\orcidlink{0000-0002-1125-7384},}
\author[34,21]{{R.~Miquel},}
\author[35]{{J.~Moustakas}\orcidlink{0000-0002-2733-4559},}
\author[24]{{S.~Nadathur}\orcidlink{0000-0001-9070-3102},}
\author[36,37]{{G.~Niz}\orcidlink{0000-0002-1544-8946},}
\author[5,4]{{N.~Palanque-Delabrouille}\orcidlink{0000-0003-3188-784X},}
\author[6,7,8]{{W.~J.~Percival}\orcidlink{0000-0002-0644-5727},}
\author[38]{{F.~Prada}\orcidlink{0000-0001-7145-8674},}
\author[39]{{I.~P\'erez-R\`afols}\orcidlink{0000-0001-6979-0125},}
\author[40]{{G.~Rossi},}
\author[41,42,43]{{L.~Samushia}\orcidlink{0000-0002-1609-5687},}
\author[44]{{E.~Sanchez}\orcidlink{0000-0002-9646-8198},}
\author[4]{{D.~Schlegel},}
\author[28,29]{{M.~Schubnell},}
\author[45]{{H.~Seo}\orcidlink{0000-0002-6588-3508},}
\author[4]{{J.~Silber}\orcidlink{0000-0002-3461-0320},}
\author[19]{{D.~Sprayberry},}
\author[29]{{G.~Tarl\'{e}}\orcidlink{0000-0003-1704-0781},}
\author[19]{{B.~A.~Weaver},}
\author[32]{{P.~Zarrouk}\orcidlink{0000-0002-7305-9578},}
\author[4]{{R.~Zhou}\orcidlink{0000-0001-5381-4372},}
\author[46]{{H.~Zou}\orcidlink{0000-0002-6684-3997},}

\affiliation[1]{Center for Cosmology and AstroParticle Physics, The Ohio State University, 191 West Woodruff Avenue, Columbus, OH 43210, USA}
\affiliation[2]{Department of Physics, The Ohio State University, 191 West Woodruff Avenue, Columbus, OH 43210, USA}
\affiliation[3]{The Ohio State University, Columbus, 43210 OH, USA}
\affiliation[4]{Lawrence Berkeley National Laboratory, 1 Cyclotron Road, Berkeley, CA 94720, USA}
\affiliation[5]{IRFU, CEA, Universit\'{e} Paris-Saclay, F-91191 Gif-sur-Yvette, France}
\affiliation[6]{Department of Physics and Astronomy, University of Waterloo, 200 University Ave W, Waterloo, ON N2L 3G1, Canada}
\affiliation[7]{Perimeter Institute for Theoretical Physics, 31 Caroline St. North, Waterloo, ON N2L 2Y5, Canada}
\affiliation[8]{Waterloo Centre for Astrophysics, University of Waterloo, 200 University Ave W, Waterloo, ON N2L 3G1, Canada}
\affiliation[9]{Department of Astronomy, The Ohio State University, 4055 McPherson Laboratory, 140 W 18th Avenue, Columbus, OH 43210, USA}
\affiliation[10]{Department of Physics, Boston University, 590 Commonwealth Avenue, Boston, MA 02215 USA}
\affiliation[11]{Department of Physics \& Astronomy, University of Rochester, 206 Bausch and Lomb Hall, P.O. Box 270171, Rochester, NY 14627-0171, USA}
\affiliation[12]{Dipartimento di Fisica ``Aldo Pontremoli'', Universit\`a degli Studi di Milano, Via Celoria 16, I-20133 Milano, Italy}
\affiliation[13]{INAF-Osservatorio Astronomico di Brera, Via Brera 28, 20122 Milano, Italy}
\affiliation[14]{Department of Physics \& Astronomy, University College London, Gower Street, London, WC1E 6BT, UK}
\affiliation[15]{Institut d'Estudis Espacials de Catalunya (IEEC), c/ Esteve Terradas 1, Edifici RDIT, Campus PMT-UPC, 08860 Castelldefels, Spain}
\affiliation[16]{Institute of Space Sciences, ICE-CSIC, Campus UAB, Carrer de Can Magrans s/n, 08913 Bellaterra, Barcelona, Spain}
\affiliation[17]{Instituto de F\'{\i}sica, Universidad Nacional Aut\'{o}noma de M\'{e}xico,  Circuito de la Investigaci\'{o}n Cient\'{\i}fica, Ciudad Universitaria, Cd. de M\'{e}xico  C.~P.~04510,  M\'{e}xico}
\affiliation[18]{Department of Astronomy, San Diego State University, 5500 Campanile Drive, San Diego, CA 92182, USA}
\affiliation[19]{NSF NOIRLab, 950 N. Cherry Ave., Tucson, AZ 85719, USA}
\affiliation[20]{University of California, Berkeley, 110 Sproul Hall \#5800 Berkeley, CA 94720, USA}
\affiliation[21]{Institut de F\'{i}sica d’Altes Energies (IFAE), The Barcelona Institute of Science and Technology, Edifici Cn, Campus UAB, 08193, Bellaterra (Barcelona), Spain}
\affiliation[22]{Departamento de F\'isica, Universidad de los Andes, Cra. 1 No. 18A-10, Edificio Ip, CP 111711, Bogot\'a, Colombia}
\affiliation[23]{Observatorio Astron\'omico, Universidad de los Andes, Cra. 1 No. 18A-10, Edificio H, CP 111711 Bogot\'a, Colombia}
\affiliation[24]{Institute of Cosmology and Gravitation, University of Portsmouth, Dennis Sciama Building, Portsmouth, PO1 3FX, UK}
\affiliation[25]{Fermi National Accelerator Laboratory, PO Box 500, Batavia, IL 60510, USA}
\affiliation[26]{Institut d'Astrophysique de Paris. 98 bis boulevard Arago. 75014 Paris, France}
\affiliation[27]{School of Mathematics and Physics, University of Queensland, Brisbane, QLD 4072, Australia}
\affiliation[28]{Department of Physics, University of Michigan, 450 Church Street, Ann Arbor, MI 48109, USA}
\affiliation[29]{University of Michigan, 500 S. State Street, Ann Arbor, MI 48109, USA}
\affiliation[30]{Department of Physics, The University of Texas at Dallas, 800 W. Campbell Rd., Richardson, TX 75080, USA}
\affiliation[31]{Department of Physics and Astronomy, University of California, Irvine, 92697, USA}
\affiliation[32]{Sorbonne Universit\'{e}, CNRS/IN2P3, Laboratoire de Physique Nucl\'{e}aire et de Hautes Energies (LPNHE), FR-75005 Paris, France}
\affiliation[33]{Departament de F\'{i}sica, Serra H\'{u}nter, Universitat Aut\`{o}noma de Barcelona, 08193 Bellaterra (Barcelona), Spain}
\affiliation[34]{Instituci\'{o} Catalana de Recerca i Estudis Avan\c{c}ats, Passeig de Llu\'{\i}s Companys, 23, 08010 Barcelona, Spain}
\affiliation[35]{Department of Physics and Astronomy, Siena University, 515 Loudon Road, Loudonville, NY 12211, USA}
\affiliation[36]{Departamento de F\'{\i}sica, DCI-Campus Le\'{o}n, Universidad de Guanajuato, Loma del Bosque 103, Le\'{o}n, Guanajuato C.~P.~37150, M\'{e}xico}
\affiliation[37]{Instituto Avanzado de Cosmolog\'{\i}a A.~C., San Marcos 11 - Atenas 202. Magdalena Contreras. Ciudad de M\'{e}xico C.~P.~10720, M\'{e}xico}
\affiliation[38]{Instituto de Astrof\'{i}sica de Andaluc\'{i}a (CSIC), Glorieta de la Astronom\'{i}a, s/n, E-18008 Granada, Spain}
\affiliation[39]{Departament de F\'isica, EEBE, Universitat Polit\`ecnica de Catalunya, c/Eduard Maristany 10, 08930 Barcelona, Spain}
\affiliation[40]{Department of Physics and Astronomy, Sejong University, 209 Neungdong-ro, Gwangjin-gu, Seoul 05006, Republic of Korea}
\affiliation[41]{Abastumani Astrophysical Observatory, Tbilisi, GE-0179, Georgia}
\affiliation[42]{Department of Physics, Kansas State University, 116 Cardwell Hall, Manhattan, KS 66506, USA}
\affiliation[43]{Faculty of Natural Sciences and Medicine, Ilia State University, 0194 Tbilisi, Georgia}
\affiliation[44]{CIEMAT, Avenida Complutense 40, E-28040 Madrid, Spain}
\affiliation[45]{Department of Physics \& Astronomy, Ohio University, 139 University Terrace, Athens, OH 45701, USA}
\affiliation[46]{National Astronomical Observatories, Chinese Academy of Sciences, A20 Datun Road, Chaoyang District, Beijing, 100101, P.~R.~China}